\documentclass[aps,prb,twocolumn,superscriptaddress,longbibliography]{revtex4-2}
\usepackage[pdftex]{graphicx}
\usepackage{mmap}
\usepackage{amssymb}
\usepackage{amsmath}
\usepackage{color}
\usepackage[usenames,dvipsnames]{xcolor}
\usepackage{hyperref}
\hypersetup{colorlinks=true,linkcolor=blue,urlcolor=blue,citecolor=blue}

\begin{document}
\title{Bose-Einstein condensation in canonical ensemble with fixed total momentum}
\author{Andrey S. Plyashechnik}%
\email{asplyashechnik@vniia.ru}%
\affiliation{N.~L. Dukhov Research Institute of Automatics (VNIIA), 127055 Moscow, Russia}%
\author{Alexey A. Sokolik}%
\email{asokolik@hse.ru}%
\affiliation{Institute for Spectroscopy RAS, 142190 Troitsk, Moscow, Russia}%
\affiliation{National Research University Higher School of Economics, 109028 Moscow, Russia}%
\author{Yurii E. Lozovik}%
\affiliation{Institute for Spectroscopy RAS, 142190 Troitsk, Moscow, Russia}%
\affiliation{National Research University Higher School of Economics, 109028 Moscow, Russia}%

\begin{abstract}
We consider Bose-Einstein condensation of noninteracting homogeneous three-dimensional gas in canonical ensemble when both particle number $N$ and total momentum $\mathbf{P}$ of all particles are fixed. Using the saddle point method, we derive the large-$N$ analytical approximations for partition function, free energy, and statistical distributions of occupation numbers of different single-particle energy levels. At temperatures below the critical point of phase transition, we predict, in some ranges of $\mathbf{P}$, fragmentation of the condensate, when more than one single-particle level is macroscopically occupied. The occupation number distributions have approximately Gaussian shapes for the levels hosting the condensate, and exponential shapes for other, noncondensate levels. Our analysis demonstrates breaking of Galilean invariance of moving finite-temperature many-particle system in the presence of Bose-Einstein condensation and extends the theory of moving and rotating quantum systems to the finite-temperature large-$N$ limit.
\end{abstract}

\maketitle

\section{Introduction}

Theory of Bose-Einstein condensation (BEC), at the simplest level of ideal gas, relies on the balance equation for a total number of particles $N$ statistically distributed among single-particle energy levels in grand canonical ensemble
\cite{Griffin,Pethick,Ziff}. For interacting Bose gas with relatively weak interaction and large condensate fraction, the Bogoliubov theory and Gross-Pitaievskii equation are the most frequently used approaches \cite{Proukakis}, and quantum Monte Carlo simulations are employed in more difficult cases of strongly interacting systems \cite{Ceperley}.

Since grand canonical ensemble becomes inappropriate and sometimes provides wrong predictions for a system of fixed number of particles $N$, the theoretical treatments starting from canonical ensemble for ideal and interacting BECs were developed (see, for example, Refs.~\cite{Kocharovsky2006,Svidzinsky,Cockburn,Kocharovsky2010,Borrmann,Wang,Weiss,Holthaus,Wang_Ma,Idziaszek,Kocharovsky_PRL} and references therein). Assuming finite and fixed $N$ is especially important when we analyze persistent or superfluid currents arising in a ring-shaped mesoscopic quantum system, because its internal energy is $N\hbar$-periodic function of angular momentum, whose minima correspond to metastable current-carrying states \cite{Byers,Bloch}. Such metastable rotating states of superfluid helium and ring-shaped atomic BECs, and their decay through phase slippage and vortex penetration were extensively studied theoretically using macroscopic hydrodynamics \cite{Iordanskii,Langer_Fisher,Langer,Saarikoski,Varoquaux} or ab initio microscopic approaches \cite{Yannouleas,Cooper,Alon}, and observed in experiments \cite{Ryu,Ramanathan,Moulder,Chien}. Analysis of nonequilibrium BEC dynamics is also greatly facilitated by the fixed-$N$ condition effectively arising in the rapid relaxation regime \cite{Shishkov_PRL,Shishkov_Quantum} in a driven-dissipative polariton system.

Our study is aimed to fill the gap between statistical approaches to BEC based on canonical ensemble with large, but fixed particle number $N$ \cite{Kocharovsky2006,Svidzinsky,Kocharovsky2010,Cockburn}, and microscopic analysis of current-carrying (or \emph{yrast}) states of finite-$N$ bosonic systems \cite{Yannouleas,Cooper,Alon} with fixed angular momentum. We consider ideal three-dimensional Bose gas at finite temperature $T$ which has both particle number $N$ and total momentum $\mathbf{P}$ fixed as external parameters. Starting from enumeration of all possible arrangements of $N$ Bose particles among single-particle energy levels, which satisfy the total momentum constraint and occur according to the Gibbs probability distribution, we use the saddle point approach to calculate thermodynamic and statistical properties of the system in the leading and next-to-leading orders of the large-$N$ limit. These properties include partition function, free energy, and statistical distributions of particle number in the condensate. We also discover fragmentation of the condensate at certain ranges of $\mathbf{P}$ and $T$, when two or more single-particle levels become macroscopically occupied in the current-carrying system.

Since we consider the system states with fixed total momentum $\mathbf{P}$ under periodic boundary conditions, cyclic motion of particles through the system boundaries is similar (up to finite-size shape effects) to rotation along circumference of a ring-shaped system. Thus our analysis presents an approximate treatment of rotating ring-shaped BEC with fixed angular momentum. We restrict ourselves to noninteracting Bose particles, so our study generalizes the canonical ensemble theory of BEC in ideal gas \cite{Kocharovsky2006,Svidzinsky,Kocharovsky2010,Cockburn,Borrmann,Wang,Weiss,Holthaus} to the setting with fully quenched fluctuations of $\mathbf{P}$. On the other hand, our analysis, thanks to the aforementioned similarity of translational motion and rotation, extends the calculations of the energies of rotating states \cite{Yannouleas,Cooper,Alon} to the large-$N$ finite-temperature limit, although for noninteracting Bose system. Note that Refs.~\cite{Tarasov2018,Cheng} studied the influence of boundary conditions on thermodynamic properties of BEC, and twisting these conditions on some nonzero phase effectively shifts the thermal distribution of $\mathbf{P}$ to a nonzero average value. In contrast, our approach assumes not just shift of average value of $\mathbf{P}$ but complete quench of its fluctuations, so it has higher generality.

The article is organized as follows. In Sec.~\ref{Sec2} we provide an overview of analytical and numerical methods which allows us to study the thermal statistics of noninteracting Bose system at fixed $N$ and $\mathbf{P}$, arriving to the saddle point approach, which is used in Sec.~\ref{Sec3} to analyze the BEC conditions and identify three distinct phases of the system: normal phase, unfragmented BEC phase, and fragmented BEC phase. In Sec.~\ref{Sec4} we calculate the free energy of the system as function of $T$, $N$, $\mathbf{P}$, and discuss the condensate fragmentation. In Sec.~\ref{Sec5} we calculate statistical distributions of occupation numbers of single-particle energy levels, which, depending on the level and system phase, has approximately Gaussian or exponential shapes. In Sec.~\ref{Sec6} we consider the role of total momentum fixation, i.e. what is the difference between our results obtained at fixed $\mathbf{P}=0$ and conventional canonical-ensemble approaches to BEC where $\mathbf{P}$ freely fluctuates. Sec.~\ref{Sec7} concludes our paper, and Appendices~\ref{Appendix_A} and \ref{Appendix_B} provide calculation details. More details on estimating the sums-over-states arising in our calculations and deriving the distribution functions are given in the Supplemental Material \cite{Supplemental}.

\section{Analytical and numerical methods}\label{Sec2}

Consider $N$ noninteracting bosons at the temperature $T=\beta^{-1}$ in a cubic box of the volume $V=L^3$ with periodic boundary conditions imposed on single-particle wave functions. The dimensionless single-particle momenta, taken in the units of the minimal momentum $k_0=2\pi\hbar/L$, are quantized integer-valued vectors  $\mathbf{k}=\{n_x,n_y,n_z\}$, and the single-particle energies are $\varepsilon_\mathbf{k}=k_0^2\mathbf{k}^2/2m$. We introduce the dimensionless total momentum
\begin{equation}
\mathbf{Q}=\mathbf{P}/k_0,
\end{equation}
which is also an integer-valued vector.

Partition function of the system with fixed particle number $N$ and total momentum $\mathbf{Q}$ reads
\begin{equation}
Z_{N\mathbf{Q}}=\sum_{\{n_\mathbf{k}\}}e^{-\beta\sum\limits_\mathbf{k}n_\mathbf{k}\varepsilon_\mathbf{k}}\delta_{N,\sum\limits_\mathbf{k}n_\mathbf{k}}\delta_{\mathbf{Q},\sum\limits_\mathbf{k}\mathbf{k}n_\mathbf{k}}.\label{Z_NQ1}
\end{equation}
Here the sum over the set of occupation numbers $n_\mathbf{k}=0,1,2,\ldots$ is restricted by the Kronecker symbols $\delta$ to fulfill the fixation conditions for $N$ and $\mathbf{Q}$. Probability to find $N_\mathbf{q}$ particles in the single-particle state with momentum $\mathbf{q}$ in our restricted ensemble is
\begin{equation}
p_{N\mathbf{Q}}(N_\mathbf{q})=\frac1{Z_{N\mathbf{Q}}}\sum_{\{n_\mathbf{k}\}}e^{-\beta\sum\limits_\mathbf{k}n_\mathbf{k}\varepsilon_\mathbf{k}}\delta_{N,\sum\limits_\mathbf{k}n_\mathbf{k}}\delta_{\mathbf{Q},\sum\limits_\mathbf{k}\mathbf{k}n_\mathbf{k}}\delta_{N_\mathbf{q},n_\mathbf{q}}.\label{p1}
\end{equation}
Note that if we shift $\mathbf{Q}$ on the vector $N\mathbf{s}$, where $\mathbf{s}=\{s_x,s_y,s_z\}$ is arbitrary integer vector, then Eqs.~(\ref{Z_NQ1}) and (\ref{p1}) transform as
\begin{align}
&Z_{N,\mathbf{Q}+N\mathbf{s}}=e^{-\frac{\beta k_0^2}{2mN}\{(\mathbf{Q}+N\mathbf{s})^2-\mathbf{Q}^2\}}Z_{N\mathbf{Q}},\label{periodicity1}\\
&p_{N,\mathbf{Q}+N\mathbf{s}}(N_{\mathbf{q}+\mathbf{s}}=n)=p_{N\mathbf{Q}}(N_\mathbf{q}=n).
\end{align}
Physically it means that the quantum state of relative motion of $N$ bosons moving on a three-dimensional torus depend on $Q_{x,y,z}$ periodically with the period $N$ \cite{Bloch}. In combination with the symmetry of $Z_{N\mathbf{Q}}$ and $p_{N\mathbf{Q}}(N_{\mathbf{q}})$ with respect to reflections $Q_j\rightarrow-Q_j$, $q_j\rightarrow-q_j$, it implies that it is sufficient to consider only the total momenta in the range $0\leqslant Q_{x,y,z}\leqslant N/2$ to probe all physically distinct states of the system.

The widely used methods \cite{Kocharovsky2006,Svidzinsky,Kocharovsky2010,Cockburn,Borrmann,Wang,Weiss} to calculate partition function of the noninteracting Bose system in canonical ensemble with fixed $N$ rely on recursion relations. In the case of fixed total momentum, we can straightforwardly derive the similar relation
\begin{equation}
Z_{N\mathbf{Q}}=\frac1N\sum_{l=1}^N\sum_\mathbf{k}e^{-l\beta\varepsilon_\mathbf{k}}Z_{N-l,\mathbf{Q}-l\mathbf{k}},\label{Z_rec}
\end{equation}
which relates $Z_{N\mathbf{Q}}$ to partition functions at other total momenta and smaller particle numbers. Additional formula can be derived to relate the probabilities (\ref{p1}) to partition functions:
\begin{align}
p_{N\mathbf{Q}}(N_\mathbf{q})&=\frac{e^{-\beta N_\mathbf{q}\varepsilon_\mathbf{q}}}{Z_{N\mathbf{Q}}}\left\{Z_{N-N_\mathbf{q},\mathbf{Q}-N_\mathbf{q}\mathbf{q}}\right.\nonumber\\
&\left.-e^{-\beta\varepsilon_\mathbf{q}}Z_{N-N_\mathbf{q}-1,\mathbf{Q}-(N_\mathbf{q}+1)\mathbf{q}}\right\}.\label{p_rec}
\end{align}
Applying the recurrence relation (\ref{Z_rec}) requires keeping in memory the array of $Z_{N\mathbf{Q}}$ of the size $\sim N^4$ and the same order of computation steps, which makes this method feasible at not very large $N$ (no more than 100 for desktop computer calculations).

Another method to calculate the restricted sums (\ref{Z_NQ1}) and (\ref{p1}) is based on the integral representation $\delta_{jl}=(2\pi i)^{-1}\int_{-\pi i}^{\pi i}dz\,e^{(j-l)z}$ of the Kronecker symbol \cite{Sinner}. Applying it in Eqs.~(\ref{Z_NQ1}) and (\ref{p1}), we obtain
\begin{align}
Z&=\int\limits_{-\pi i}^{\pi i}\frac{dzd\mathbf{r}}{(2\pi i)^4}e^{-zN-\mathbf{r}\cdot\mathbf{Q}}\sum_{\{n_\mathbf{k}\}}e^{\sum\limits_\mathbf{k}(-\beta\varepsilon_\mathbf{k}+z+\mathbf{r}\cdot\mathbf{k})n_\mathbf{k}},\label{Z_NQ2}\\
p(N_\mathbf{q})&=\int\limits_{-\pi i}^{\pi i}\frac{dzd\mathbf{r}}{(2\pi i)^4Z}e^{-z(N-N_\mathbf{q})-\mathbf{r}\cdot(\mathbf{Q}-\mathbf{q}N_\mathbf{q})}\nonumber\\
&\times e^{-\beta N_\mathbf{q}\varepsilon_\mathbf{q}}\sum_{\{n_{\mathbf{k}\neq\mathbf{q}}\}}e^{\sum\limits_\mathbf{k}(-\beta\varepsilon_\mathbf{k}+z+\mathbf{r}\cdot\mathbf{k})n_\mathbf{k}}.\label{p2}
\end{align}
Here and in the rest of the paper, we omit the indices $N$, $\mathbf{Q}$ of $Z$ and $p$ to avoid clutter. It is convenient to introduce the dimensionless parameter
\begin{equation}
R=\frac{2m}{\beta k_0^2}=\frac{L^2}{\pi\lambda_\mathrm{th}^2}=\frac{N^{2/3}}{\pi[\zeta(\frac32)]^{2/3}}\frac{T}{T_\mathrm{c}},\label{R}
\end{equation}
where $\zeta(x)$ is the Riemann zeta function, $\lambda_\mathrm{th}=\hbar\sqrt{2\pi\beta/m}$ is the thermal de Broglie wavelength, and
\begin{equation}
T_\mathrm{c}=\frac{2\pi}{[\zeta(\frac32)]^{2/3}}\frac{\hbar^2}m\left(\frac{N}{L^3}\right)^{2/3}\approx3.31\frac{\hbar^2}m\left(\frac{N}{L^3}\right)^{2/3}\label{T_c}
\end{equation}
is the conventional critical temperature of BEC in the thermodynamic limit. Changing the variable $\mathbf{r}=2\mathbf{v}/R$ in Eqs.~(\ref{Z_NQ2})--(\ref{p2}), using relation $\beta\varepsilon_\mathbf{k}=\mathbf{k}^2/R$, and performing summations over $n_\mathbf{k}$ which are now unrestricted, we obtain
\begin{align}
Z=&\frac1{2\pi^4R^3}\int\limits_{-\pi i}^{\pi i}dz\int\limits_{-\pi iR/2}^{\pi iR/2}d\mathbf{v}\:e^{f(z,\mathbf{v})},\label{Z_NQ3}\\
p(N_\mathbf{q})=&\frac1{2\pi^4R^3Z}\int\limits_{-\pi i}^{\pi i}dz\int\limits_{-\pi iR/2}^{\pi iR/2}d\mathbf{v}\:e^{\tilde{f}(N_\mathbf{q},z,\mathbf{v})},\label{p3}
\end{align}
where
\begin{equation}
f(z,\mathbf{v})=-zN-\frac{2\mathbf{v}\cdot\mathbf{Q}}R-\sum_\mathbf{k}\log\left(1-e^{\frac{-\mathbf{k}^2+2\mathbf{v}\cdot\mathbf{k}}R+z}\right),\label{f1}
\end{equation}
\begin{align}
\tilde{f}(N_\mathbf{q},z,\mathbf{v})&=-z(N-N_\mathbf{q})-\frac{2\mathbf{v}\cdot(\mathbf{Q}-\mathbf{q}N_\mathbf{q})}R\nonumber\\
&-\frac{\mathbf{q}^2N_\mathbf{q}}R-\sum_{\mathbf{k}\neq\mathbf{q}}\log\left(1-e^{\frac{-\mathbf{k}^2+2\mathbf{v}\cdot\mathbf{k}}R+z}\right).\label{g1}
\end{align}

\section{Saddle-point calculation of partition function}\label{Sec3}

\begin{figure}[t]
\begin{center}
\includegraphics[width=\columnwidth]{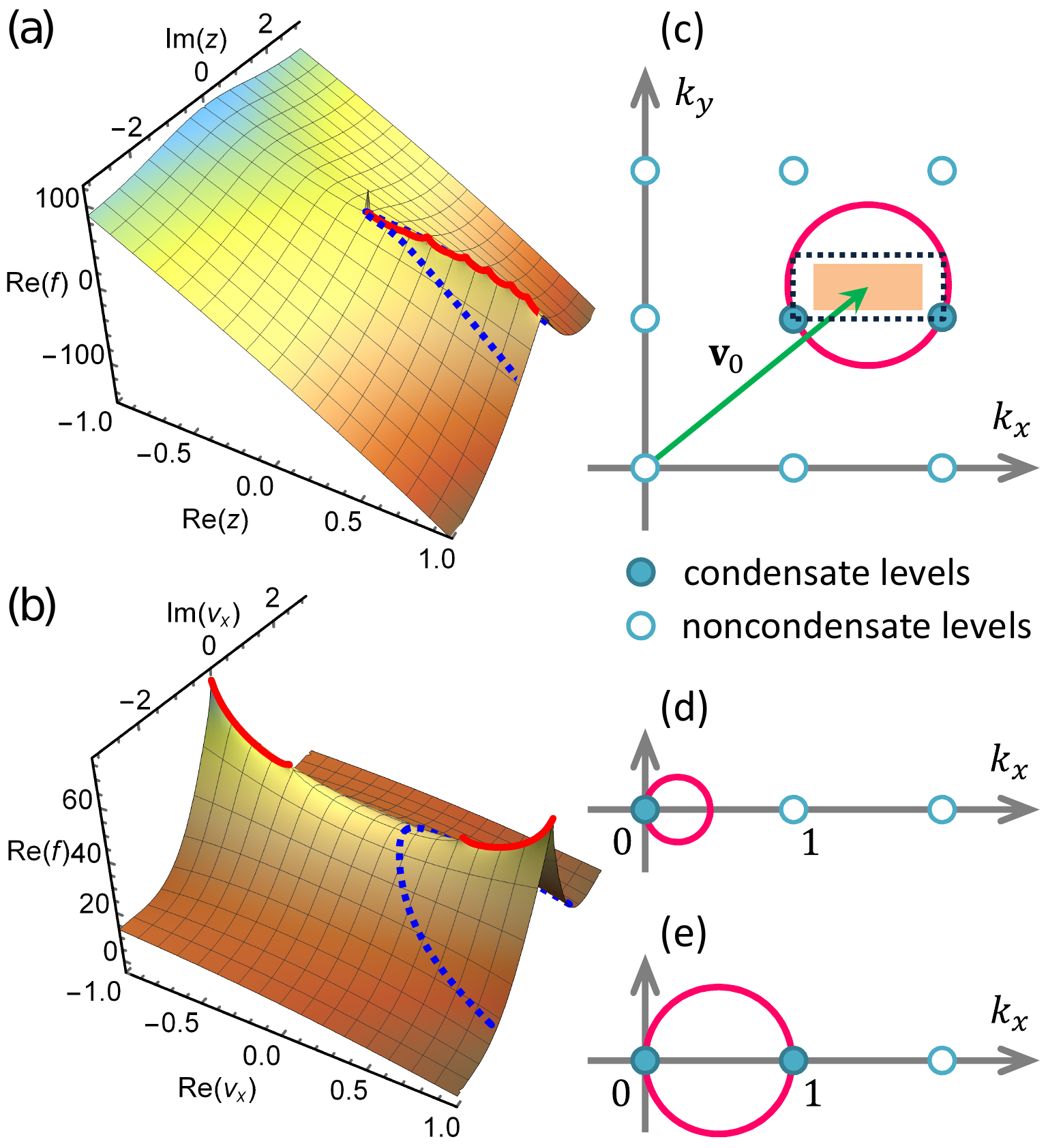}
\end{center}
\caption{\label{Fig1} (a,b) Real part of the function $f(z,\mathbf{v})$ and the steepest-descent paths (blue dotted lines) passing through the saddle point $(z_0,\mathbf{v}_0)$: (a) on the complex plane of $z$ at $\mathbf{v}=\mathbf{v}_0$, (b) on the complex plane of $v_x$ at $z=z_0$, $v_{y,z}=0$. Red solid lines show branch cuts of $f$, and calculation parameters are $N=100$, $T=0.8T_\mathrm{c}$, $\mathbf{Q}=\{10,0,0\}$. (c) Two-dimensional space with quantized single-particle momenta $\mathbf{k}$, two of them (filled circles) satisfy Eq.~(\ref{cond}) and host the condensate, while the remaining ones (empty circles) comprise the thermal cloud. Shaded rectangle depicts allowable region for the total momenta per particle $\mathbf{Q}/N$ restricted by inequality (\ref{Q_ineq}). (d,e) Similar pictures showing unfragmented (at $0\leqslant v_{0x}<1/2$) and fragmented (at $v_{0x}=1/2$) BEC states, respectively, at $\mathbf{Q}=\mathbf{e}_xQ$.}
\end{figure}

We will calculate the integrals (\ref{Z_NQ3})--(\ref{p3}) in the large-$N$ limit with a help of the saddle point approximation, as usually done when switching between different statistical ensembles \cite{Huang}. The saddle point $(z_0,\mathbf{v}_0)$ for the integral (\ref{Z_NQ3}) is obtained by equating the derivatives of the exponent (\ref{f1}) to zero:
\begin{align}
\frac{\partial f}{\partial z}&=-N+\sum_\mathbf{k}\nu_\mathbf{k}=0,\label{fder1}\\
\frac{\partial f}{\partial\mathbf{v}}&=\frac2R\left(-\mathbf{Q}+\sum_\mathbf{k}\mathbf{k}\nu_\mathbf{k}\right)=0,\label{fder2}
\end{align}
where
\begin{equation}
\nu_\mathbf{k}=\frac1{e^{\frac{\mathbf{k}^2-2\mathbf{v}_0\cdot\mathbf{k}}R-z_0}-1}\label{nu_k}
\end{equation}
is the Bose-Einstein distribution for particles in the frame moving with velocity $k_0\mathbf{v}_0/m$, where the chemical potential is $\mu=Tz_0+k_0^2\mathbf{v}_0^2/2m$. An example of the steepest-descent trajectories passing through the saddle point is shown in Fig.~\ref{Fig1}(a,b). Physically, Eqs. (\ref{fder1})--(\ref{fder2}) impose the balance conditions on the total particle number $N$ and the total momentum $\mathbf{Q}$ in the moving reference frame. Therefore we can treat the quantity $\mathbf{v}_0$, called hereafter \emph{rapidity}, as dimensionless velocity of the frame where the average momentum vanishes, and $z_0$ as dimensionless effective chemical potential (apart from the Galilean transformation term $\propto\mathbf{v}_0^2$) of noninteracting bosons in this frame. 

Note that the saddle-point method is applicable when the function $f(z,\mathrm{v})$ is not too strongly skewed near the saddle point $(z_0,\mathbf{v}_0)$, i.e. its Taylor expansion near this point can be safely cut beyond quadratic terms. Estimating 2nd and 3rd derivatives of $f$, we find that this condition is violated when $R\lesssim1$. Therefore $R\gg1$, or $T/T_\mathrm{c}\gg N^{-2/3}$ is the necessary condition for applicability of the saddle point method, although numerical validation demonstrates its good accuracy already at $R>2$. In the following, we will assume in our calculations that $T/T_\mathrm{c}=\mathrm{const}$ when we consider the thermodynamic limit $N\rightarrow\infty$, so that $R\sim N^{2/3}\rightarrow\infty$ in this limit. In the low-temperature regime, when $T/T_\mathrm{c}$ is constant but much less than 1, we need large enough $N$ to make sure that $R\gg1$.

Occupation numbers (\ref{nu_k}) are finite when 
\begin{equation}
(\mathbf{k}-\mathbf{v}_0)^2>\mathbf{v}_0^2+z_0R\label{noncond}
\end{equation}
for any quantized momentum $\mathbf{k}$. One or several $\mathbf{k}$-levels can host the condensate, $\nu_\mathbf{k}\propto N$, when $(\mathbf{k}-\mathbf{v}_0)^2-\mathbf{v}_0^2-z_0R=\mathcal{O}(N^{-1/3})$ in the limit $N\rightarrow\infty$, i.e. when the inequality (\ref{noncond}) turns into equality
\begin{equation}
(\mathbf{k}-\mathbf{v}_0)^2=\mathbf{v}_0^2+z_0R\label{cond}
\end{equation}
in the thermodynamic limit (note that the chemical potential $z_0$ is not necessary of the order of $N^{-1}$ as in conventional BEC picture and can be of the order of $N^{-2/3}$ at large enough $|\mathbf{Q}|$ due to Galilean shift). As demonstrated in Fig.~\ref{Fig1}(c) for simplified two-dimensional picture, Eq.~(\ref{cond}) is the condition for a sphere of the maximum allowed radius $\sqrt{\mathbf{v}_0^2+z_0R}$ centered at $\mathbf{v}_0$, which touches the nearest discrete $\mathbf{k}$-points that become the single-particle \emph{condensate levels}. We denote the set of such levels (up to 8 in three dimensions) by $C$. Other levels outside this sphere satisfying Eq.~(\ref{noncond}) are \emph{noncondensate} ones and comprise the thermal cloud.

As shown in Appendix~\ref{Appendix_A}, the contributions of the thermal cloud to the total particle number and momentum can be approximated using the Bose-Einstein integral $g_d(x)$ with a subextensive error, so in the leading $\mathcal{O}(N)$ order the saddle-point conditions (\ref{fder1})--(\ref{fder2}) read:
\begin{align}
\sum_{\mathbf{k}\in C}\nu_\mathbf{k}+\frac{N}{\zeta(\frac32)}\left(\frac{T}{T_\mathrm{c}}\right)^{3/2}g_{3/2}\left(-z_0-\frac{\mathbf{v}_0^2}R\right)=N,\label{fder3}\\
\sum_{\mathbf{k}\in C}\mathbf{k}\nu_\mathbf{k}+\frac{\mathbf{v}_0N}{\zeta(\frac32)}\left(\frac{T}{T_\mathrm{c}}\right)^{3/2}g_{3/2}\left(-z_0-\frac{\mathbf{v}_0^2}R\right)=\mathbf{Q},\label{fder4}
\end{align}
with the omitted subleading corrections of the order $N^{2/3}$. Since $g_{3/2}(-z_0-\mathbf{v}_0^2/R)$ reaches the maximal value $\zeta(\frac32)$ when its argument vanishes, the last term in the left-hand side of Eq.~(\ref{fder3}) is bounded by $N(T/T_\mathrm{c})^{3/2}$ from above. Therefore at $T>T_\mathrm{c}$ this equation can be satisfied at $-z_0-\mathbf{v}_0^2/R\sim1$ in the limit $N\rightarrow\infty$. It is the normal-state regime with no condensate levels, where Eqs.~(\ref{fder3})--(\ref{fder4}) reduce to equations for chemical potential
\begin{equation}
\frac1{\zeta(\frac32)}\left(\frac{T}{T_\mathrm{c}}\right)^{3/2}g_{3/2}\left(-z_0-\frac{\mathbf{v}_0^2}R\right)=1\label{z_norm}
\end{equation}
and rapidity
\begin{equation}
\mathbf{v}_0\approx\frac{\mathbf{Q}}N=\frac{\mathbf{P}}{mN}.\label{v_norm}
\end{equation}
Eq.~(\ref{v_norm}) states that the normal Bose gas obeys Galilean invariance since the occupation numbers (\ref{nu_k}) turn out to be simply displaced in the $\mathbf{k}$-space due to common center-of-mass motion.

On the contrary, at $T<T_\mathrm{c}$ the Galilean invariance is broken: the noncondensate term in Eq.~(\ref{fder3}) is saturated at $z_0\rightarrow-\mathbf{v}_0^2/R$ and cannot accommodate all $N$ particles. In this limit, Eqs.~(\ref{fder3})--(\ref{fder4}) in the leading $\mathcal{O}(N)$ order take the form
\begin{align}
\sum_{\mathbf{k}\in C}\nu_\mathbf{k}+N\left(\frac{T}{T_\mathrm{c}}\right)^{3/2}&=N,\label{N_balance1}\\
\sum_{\mathbf{k}\in C}\mathbf{k}\nu_\mathbf{k}+\mathbf{v}_0N\left(\frac{T}{T_\mathrm{c}}\right)^{3/2}&=\mathbf{Q}.\label{Q_balance1}
\end{align}
From these equations, we obtain $\sum_{\mathbf{k}\in C}(\mathbf{k}-\mathbf{v}_0)\nu_\mathbf{k}=\mathbf{Q}-\mathbf{v}_0N$ and $\sum_{\mathbf{k}\in C}\nu_\mathbf{k}=N[1-(T/T_\mathrm{c})^{3/2}]$, which allows to derive the inequality
\begin{equation}
\left|\frac{Q_i}N-v_{0i}\right|\leqslant\left[1-\left(\frac{T}{T_\mathrm{c}}\right)^{3/2}\right]\times|k_i-v_{0i}|_{\mathbf{k}\in C},\label{Q_ineq}
\end{equation}
where the distance $|k_i-v_{0i}|$ is the same for all condensate levels $\mathbf{k}\in C$. This inequality is, however, saturated and turns into equality when only a single condensate level is present. The geometrical meaning of Eq.~(\ref{Q_ineq}) is demonstrated in Fig.~\ref{Fig1}(c): if some levels host the condensate, then the weaker version of this inequality $|Q_i/N-v_{0i}|\leqslant|k_i-v_{0i}|_{\mathbf{k}\in C}$ restricts the total momentum per particle $\mathbf{Q}/N$ to the dotted rectangle, while the original Eq.~(\ref{Q_ineq}) restricts it to the shaded rectangle which is smaller by the factor $1-(T/T_\mathrm{c})^{3/2}$. This analysis helps to figure out qualitative relation between $\mathbf{Q}$ and $\mathbf{v}_0$: if the total momentum per particle $\mathbf{Q}/N$ is located in some cubic cell bounded by integer-valued points $\mathbf{k}$, the rapidity $\mathbf{v}_0$ ends up in the same cell, so generally, when $\mathbf{Q}$ is increased, $\mathbf{v}_0$ increases too. However, at $T<T_\mathrm{c}$ the strict proportionality (\ref{v_norm}) between these vectors is absent witnessing breaking of Galilean invariance.

In the forthcoming sections, for simplicity, we consider only total momenta $\mathbf{Q}$ directed along the $x$ axis, so in the following we imply $\mathbf{Q}=\mathbf{e}_xQ$. From Eq.~(\ref{fder2}) we immediately see that the saddle point
$\mathbf{v}_0$ should also lie on the $x$ axis. The physically nonequivalent momenta, as discussed above, lie in the range $0\leqslant Q\leqslant N/2$, which, due to Eq.~(\ref{Q_ineq}) corresponds to the range $0\leqslant v_{0x}\leqslant1/2$ of rapidities. In this range, we can expect one of two possibilities shown in Fig.~\ref{Fig1}(d,e). The first one is $0\leqslant v_{0x}<1/2$ [Fig.~\ref{Fig1}(d)], where only the $\mathbf{k}=0$ level (hereafter \emph{0th level}) hosts the condensate. In this case the inequality (\ref{Q_ineq}) is saturated and we obtain condition $Q<Q_0$ for the total momentum, where the threshold momentum equals
\begin{equation}
Q_0=\frac{N}2\left(\frac{T}{T_\mathrm{c}}\right)^{3/2}.\label{Q0}
\end{equation}
In this case mean occupation of the 0th condensate level is $\nu_0\approx-1/z_0$, and from the saddle-point conditions (\ref{fder3})--(\ref{fder4}) we obtain
\begin{align}
z_0&=-\frac1{N[1-(T/T_\mathrm{c})^{3/2}]}+\mathcal{O}(N^{-4/3}),\label{z_cond1}\\
v_{0x}&=\frac{Q}{2Q_0}+\mathcal{O}(N^{-1/3}),\label{v_cond1}\\
\nu_0&=N\left[1-\left(\frac{T}{T_\mathrm{c}}\right)^{3/2}\right]+\mathcal{O}(N^{2/3}),\label{nu0_cond1}\\
\nu_1&=\mathcal{O}(N^{2/3}).\label{nu1_cond1}
\end{align}
The second case is $v_{0x}=1/2$ [Fig.~\ref{Fig1}(e)], where the condensate is present in both $\mathbf{k}=0$ and $\mathbf{k}=\mathbf{e}_x$ (hereafter \emph{1st level}) simultaneously, i.e. is fragmented in the momentum space. The inequality (\ref{Q_ineq}) in this case restricts the total momentum to $Q_0<Q\leqslant N/2$. Using $\nu_0\approx-1/z_0$, $\nu_1\approx[(1-2v_{0x})/R-z_0]^{-1}$, we find from the saddle-point conditions (\ref{fder3})--(\ref{fder4})
\begin{align}
z_0&=-\frac1{N[1-\frac12(T/T_\mathrm{c})^{3/2}]-Q}+\mathcal{O}(N^{-4/3}),\label{z_cond2}\\
v_{0x}&=1/2-\mathcal{O}(N^{-1/3}),\label{v_cond2}\\
\nu_0&=N\left[1-\left(\frac{T}{T_\mathrm{c}}\right)^{3/2}\right]-(Q-Q_0)+\mathcal{O}(N^{2/3}),\label{nu0_cond2}\\
\nu_1&=Q-Q_0+\mathcal{O}(N^{2/3}).\label{nu1_cond2}
\end{align}

To conclude this section, we identify the following three qualitatively different regimes in the large-$N$ limit.

1) $T>T_\mathrm{c}$: \emph{normal phase} where BEC is absent, and the system is Galilean invariant. The saddle point location is given by Eqs.~(\ref{z_norm})--(\ref{v_norm}).

2) $T<T_\mathrm{c}$, $Q<Q_0$: \emph{unfragmented BEC phase}, where only the 0th level is macroscopically occupied. Moreover, its occupation  $\nu_0$ remains the same as in the system at rest or without momentum fixation, so BEC on the 0th level possesses some degree of robustness against total motion of the system. The saddle point location in this regime is given by Eqs.~(\ref{z_cond1})--(\ref{v_cond1}).

3) $T<T_\mathrm{c}$, $Q_0<Q\leqslant N/2$: \emph{fragmented BEC phase}, where both 0th and 1st levels are macroscopically occupied. On increase of $Q$, the occupation $\nu_0$ of the 0th level gradually decreases, and the occupation $\nu_1$ of the 1st level increases, while their sum remains the same as the standard condensate population $N[1-(T/T_\mathrm{c})^{3/2}]$ in a system without momentum fixation [see Fig.~\ref{Fig3}(b)]. The saddle point location in this regime is given by Eqs.~(\ref{z_cond2})--(\ref{v_cond2}).

\begin{figure}[t]
\begin{center}
\includegraphics[width=\columnwidth]{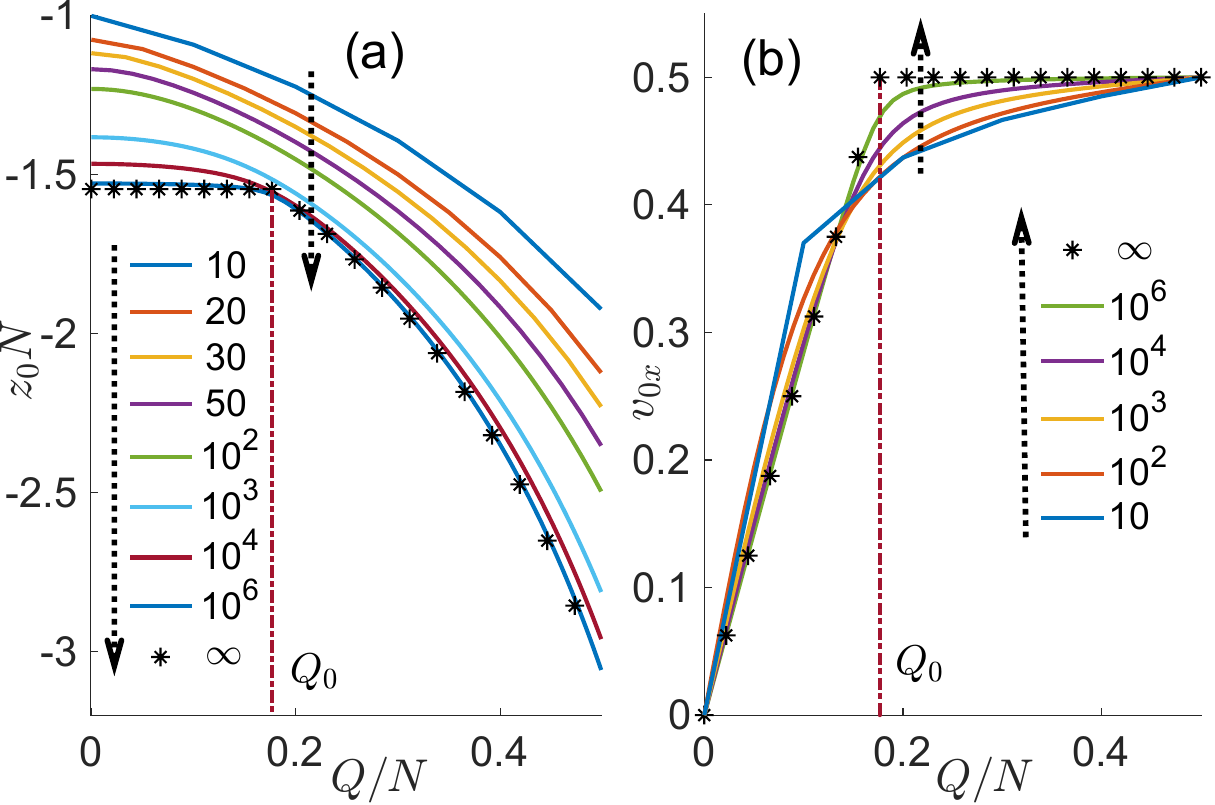}
\end{center}
\caption{\label{Fig2}Saddle point coordinates (a) $z_0$ (or dimensionless chemical potential) and (b) $v_{0x}$ (rapidity, or dimensionless mean velocity of the Bose gas) as functions of the total momentum $Q$ calculated at $T=0.5T_\mathrm{c}$ and different particle numbers $N$ shown in legends. Arrows show the order of increasing $N$. Vertical dash-dotted lines indicate the threshold momentum $Q_0$ for the BEC fragmentation. Stars show analytical approximations (\ref{z_cond1})--(\ref{v_cond1}) and (\ref{z_cond2})--(\ref{v_cond2}) obtained in the limit $N\rightarrow\infty$.}
\end{figure}

Numerical solutions of the saddle-point equations (\ref{fder1})--(\ref{fder2}) are shown in Fig.~\ref{Fig2} in the BEC region $T=0.5T_\mathrm{c}$ at different particle numbers $N$. In the limit $N\rightarrow\infty$ (stars), in the unfragmented BEC state ($Q<Q_0$) the chemical potential $\mu=Tz_0$ is constant, and the rapidity $v_{0x}$ is linearly increasing with $Q$. In the fragmented BEC state $Q>Q_0$, the chemical potential decreases, and the rapidity levels off at $v_{0x}\approx1/2$. At finite particle number $N$, these tendencies remain qualitatively the same, but the curves become smoothed, as typically happens with phase transitions in finite systems.

Going beyond the restriction $0\leqslant Q\leqslant N/2$, we can use the periodicity condition (\ref{periodicity1}) and evenness of the statistics with respect to $Q$ to draw the phase diagram shown in Fig.~\ref{Fig3}(a). At $T<T_\mathrm{c}$, the system passes through the sequence of unfragmented and fragmented BECs when its total momentum is changed. If $Q$ traverses the period $N$, all occupation numbers of the single-particle states $\mathbf{k}$ become merely shifted by $\Delta\mathbf{k}=\pm\mathbf{e}_x$ in the momentum space. The occupations of condensate levels shown in Fig.~\ref{Fig3}(b) also alternate as functions of $Q$ showing the interleaved regions of unfragmented and fragmented BECs. What happens at arbitrary directions of $\mathbf{Q}$ was described above by Eq.~(\ref{Q_ineq}) and Fig.~\ref{Fig1}(c).

\begin{figure}[t]
\begin{center}
\includegraphics[width=\columnwidth]{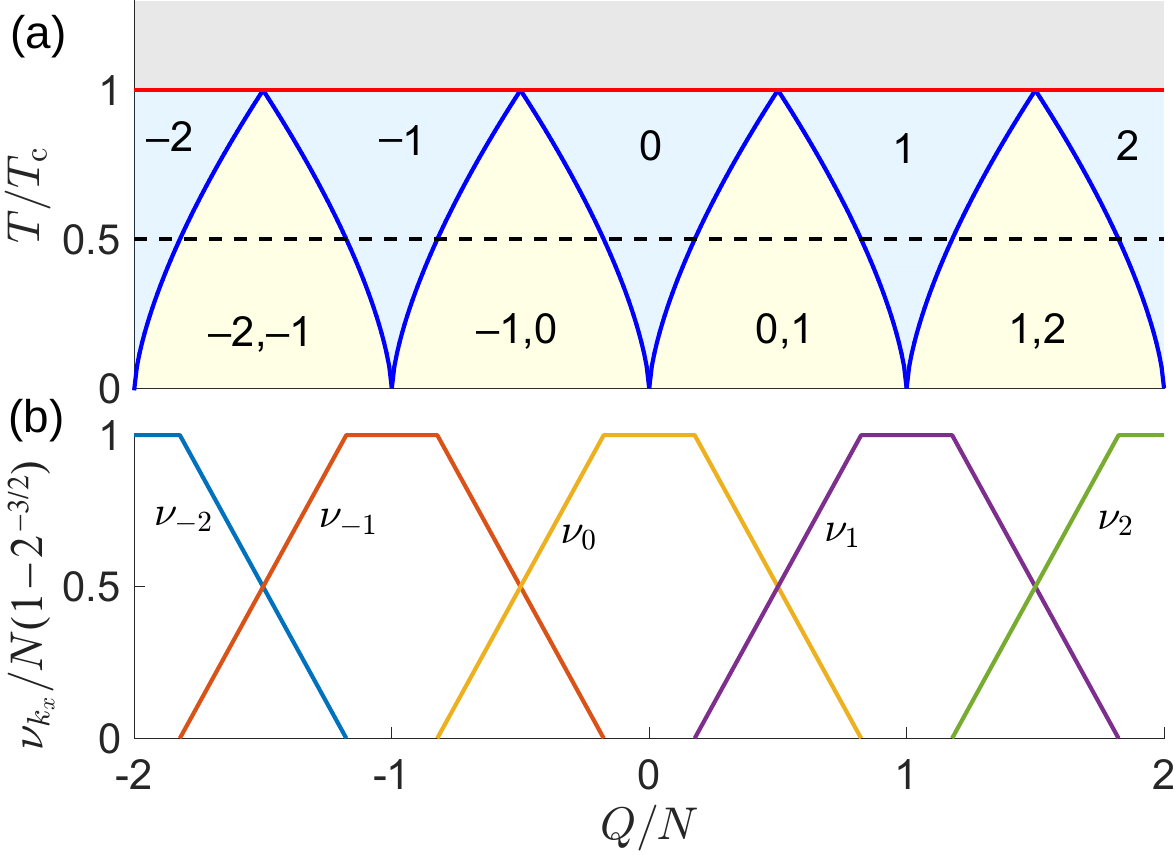}
\end{center}
\caption{\label{Fig3} (a) Phase diagram of the Bose gas in terms of its temperature $T$ and total momentum $Q$. The normal phase at $T/T_\mathrm{c}>1$ is separated from the BEC phase $T/T_\mathrm{c}<1$. In the BEC phase, the condensate can populate one or two single-particle states with momenta $k_x$ shown in the regions separated by the lines $Q\mod N=\pm Q_0$ of fragmentation transitions. (b) Mean populations $\nu_{k_x}$ of the single-particle states with momenta $k_x$ shown at $T=0.5T_\mathrm{c}$ (along the dashed line of the upper panel) in the units of the maximal condensate population $N[1-(T/T_\mathrm{c})^{3/2}]=N(1-2^{-3/2})$ as functions of the total momentum $Q$ in the limit $N\rightarrow\infty$.}
\end{figure}

\section{Thermodynamics}\label{Sec4}
\subsection{Free energy}

In the large-$N$ limit, the integral (\ref{Z_NQ3}) is dominated by vicinity of the saddle point $(z_0,\mathbf{v}_0)$. In the leading exponential order it is equal to $Z\sim e^{f(z_0,\mathbf{v}_0)}$. Using the approximate expression (\ref{log_sum}) for the logarithmic sum, we obtain the free energy $F=-T\log Z=-Tf(z_0,\mathbf{v}_0)$ as
\begin{align}
F&=NT\left\{z_0+\frac{2v_{0x}Q}{NR}-\frac1{\zeta(\frac32)}\left(\frac{T}{T_\mathrm{c}}\right)^{3/2}\right.\nonumber\\
&\times\left.g_{5/2}\left(-z_0-\frac{v_{0x}^2}R\right)\right\}+\mathcal{O}(\log N).
\end{align}
In the normal phase $T>T_\mathrm{c}$, using Eqs.~(\ref{R}), (\ref{z_norm})--(\ref{v_norm}), and performing Taylor expansion of $g_{5/2}$, we obtain in the leading order:
\begin{equation}
F\approx NT\left\{z_0-\frac1{\zeta(\frac32)}\left(\frac{T}{T_\mathrm{c}}\right)^{3/2}g_{5/2}\left(-z_0\right)\right\}+\frac{k_0^2Q^2}{2mN}.\label{F_norm}
\end{equation}
The first term is the ordinary free energy of normal noninteracting Bose gas without momentum fixation \cite{Huang}, and the second term is $\mathbf{P}^2/2mN$, i.e. the kinetic energy of entire gas moving with the total momentum $\mathbf{P}$. We again see manifestation of Galilean invariance of a normal Bose gas at $T>T_\mathrm{c}$.

\begin{figure}[t]
\begin{center}
\includegraphics[width=0.9\columnwidth]{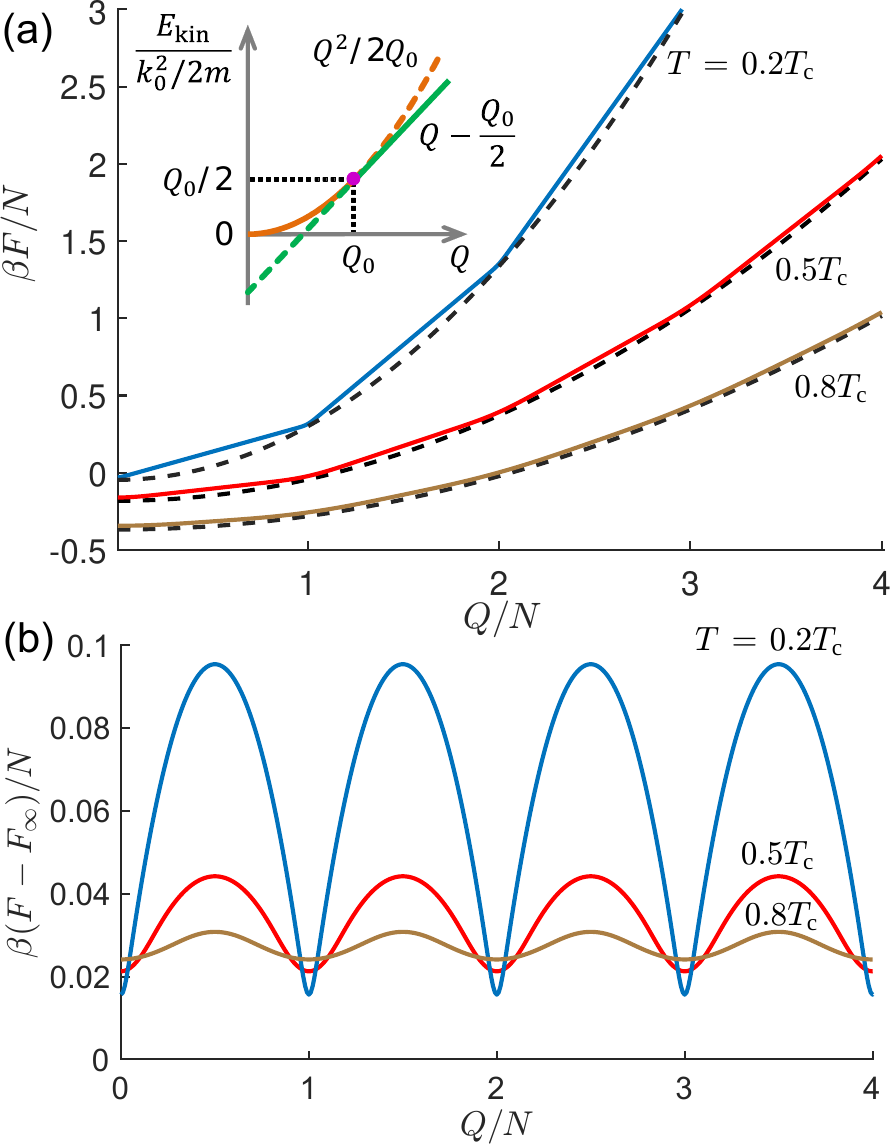}
\end{center}
\caption{\label{Fig4} (a) Free energy $F$ as function of the total momentum $Q$ at different temperatures for $N=800$. Dashed lines show the thermodynamic limits $F_\infty$ given by Eq.~(\ref{F_infty}). (b) Corresponding differences of $F$ and $F_\infty$. Inset in (a) shows the large-$N$ limit of kinetic energy $E_\mathrm{kin}$ when the momentum $Q$ is carried solely be the thermal cloud (orange parabola) or shared between the 1st level and the thermal cloud in the most favorable proportion (green straight line). The solid lines show physical branches which minimize $E_\mathrm{kin}$.}
\end{figure}

At $T<T_\mathrm{c}$ the Galilean invariance is broken, as shown in Fig.~\ref{Fig4}(a) by several examples of numerically calculated $F$ as function of $Q$. This dependence does not resemble the quadratic center-of-mass kinetic energy $k_0^2Q^2/2mN$ with the addition of a constant internal free energy, because the latter also becomes $Q$-dependent. Analytically, in the unfragmented BEC phase $T<T_\mathrm{c}$, $Q<Q_0$, using Eqs.~(\ref{R}), (\ref{z_cond1})--(\ref{v_cond1}), we obtain
\begin{equation}
F\approx-NT\frac{\zeta(\frac52)}{\zeta(\frac32)}\left(\frac{T}{T_\mathrm{c}}\right)^{3/2}+\frac{k_0^2Q^2}{2mN'}.\label{F_BEC1}
\end{equation}
Here the first term is the free energy of particles in excited $\mathbf{k}\neq0$ states (thermal cloud), which is the same as in the Bose-condensed gas without momentum fixation \cite{Huang}. The second term can be interpreted as kinetic energy $\mathbf{P}^2/2mN'$ of the thermal cloud. The mean particle number in the thermal cloud is [see Eqs.~(\ref{nu0_cond1})--(\ref{nu1_cond1})]
\begin{equation}
N'=N-\nu_0-\nu_1=N\left(\frac{T}{T_\mathrm{c}}\right)^{3/2}.\label{Np}
\end{equation}
Thus the whole momentum $\mathbf{P}$ is carried only by the thermal cloud, since the condensate in the 0th state has zero momentum.

In the fragmented BEC phase $T<T_\mathrm{c}$, $Q_0<Q\leqslant N/2$, the thermal cloud population (\ref{Np}) remains the same, and, using Eqs.~(\ref{R}), (\ref{z_cond2})--(\ref{v_cond2}), we obtain
\begin{equation}
F\approx-NT\frac{\zeta(\frac52)}{\zeta(\frac32)}\left(\frac{T}{T_\mathrm{c}}\right)^{3/2}+\frac{(Q-Q_0)k_0^2}{2m}+\frac{k_0^2Q_0^2}{2mN'}.\label{F_BEC2}
\end{equation}
The first term is the same internal free energy of the thermal cloud as in the Bose gas without momentum fixation. The second term is the kinetic energy $\nu_1k_0^2/2m$ of condensate at the 1st level, since $\nu_1=Q-Q_0$ (\ref{nu1_cond2}), and the third term is the kinetic energy of thermal cloud moving with momentum $Q_0$. Unlike the previous case of unfragmented BEC, here the total momentum $Q$ is shared between the thermal cloud and the 1st state, because it allows to lower the total kinetic energy, as discussed in the next subsection.

In agreement with Eqs.~(\ref{F_BEC1}) and (\ref{F_BEC2}), the numerically calculated $F(Q)$ at $T<T_\mathrm{c}$ [Fig.~\ref{Fig4}(a)] consists of interchanging parabolic parts in the unfragmented BEC phases near $Q=sN$ and almost linear pieces in the fragmented BEC phases in between. It conforms to the piecewise linear dependence expected at $T=0$ \cite{Bloch} but smoothed in our case due to nonzero temperatures. In Fig.~\ref{Fig4}(b) we subtract from $F$ the smooth parabolic function 
\begin{equation}
F_\infty=-NT\frac{\zeta(\frac52)}{\zeta(\frac32)}\left(\frac{T}{T_\mathrm{c}}\right)^{3/2}+\frac{k_0^2Q^2}{2mN}\label{F_infty}
\end{equation}
reached in the thermodynamic limit to obtain the extra energy of relative motion whose $Q$-dependence stems from Galilean invariance breaking. At $T=0$ this quantity is expected to have the shape of periodically repeating upturned parabolas \cite{Bloch}, but in our case junctions of these parabolas are smoothed due to nonzero temperatures.

\subsection{Fragmentation of the condensate}

To explain the fragmentation, we can write the balance equation for total momentum $Q$ and kinetic energy $E_\mathrm{kin}$ shared between the 1st level and thermal cloud,
\begin{equation}
\left\{\begin{array}{l}Q=\nu_1+Q',\\\displaystyle E_\mathrm{kin}=\nu_1\frac{k_0^2}{2m}+\frac{k_0^2Q'^2}{2mN'},\end{array}\right.
\end{equation}
where $Q'$ and $N'=2Q_0$ are, respectively, momentum and particle number of the thermal cloud. From these equations we see that the minimum of $E_\mathrm{kin}$ equal to $(k_0^2/2m)(Q-Q_0/2)$ is formally achieved at $\nu_1=Q-Q_0$, as shown in the inset in Fig.~\ref{Fig4}(a) by green straight line. However, at $Q<Q_0$ this minimum is unphysical because $\nu_1$ cannot be negative, so the physical minimum is $\nu_1=0$, $E_\mathrm{kin}=(k_0^2/2m)(Q^2/2Q_0)$ [inset in Fig.~\ref{Fig4}(a), orange parabola]. Thus the fragmentation of BEC is governed by competition between two ways to share the total momentum $Q$ between the condensate and the thermal cloud. At $Q<Q_0$ the whole momentum is carried by the thermal cloud, and transferring it to the condensate at the 1st level can only increase $E_\mathrm{kin}$. At $Q>Q_0$ transferring the momentum $Q-Q_0$ to the 1st condensate level and keeping the remaining momentum $Q'=Q_0$ in the thermal cloud allows to reduce $E_\mathrm{kin}$ and hence the free energy (since the entropic part of the free energy carried only by the thermal cloud is independent on distribution of the total momentum).

In the presence of repulsive interparticle interaction, the condensate fragmentation is usually suppressed by extra energy cost associated with exchange energy \cite{Nozieres}. Although our analysis is focused on noninteracting Bose gas, we can estimate the exchange energy hindering the fragmentation for a three-dimensional homogeneous system. Consider, following Ref.~\cite{Nozieres}, the Hamiltonian of effective contact interaction
\begin{equation}
H_\mathrm{int}=\frac{U}{2L^3}\sum_{\mathbf{k}_1\mathbf{k}_2\mathbf{q}}a^\dag_{\mathbf{k}_1+\mathbf{q}}a^\dag_{\mathbf{k}_2-\mathbf{q}}a_{\mathbf{k}_2}a_{\mathbf{k}_1},
\end{equation}
where $U$ is the interaction constant, $a_\mathbf{k}$ and $a^\dag_\mathbf{k}$ are the destruction and creation operators for Bose particles. In the mean-field approximation, where only the inter-condensate interaction is taken into account, we assume $a_\mathbf{k}\approx a^\dag_\mathbf{k}\approx \sqrt{N_\mathbf{k}}$ whenever $\mathbf{k}$ belongs to one of the condensate levels. In the presence of two condensates with populations $N_0$ and $N_1$, in the leading order we obtain
\begin{equation}
H_\mathrm{int}\approx\frac{U(N_0+N_1)^2}{2L^3}+\frac{UN_0N_1}{L^3}.
\end{equation}
The first term, being the Hartree interaction energy, does not depend on distribution of the condensate among the 0th and 1st levels. The second term, corresponding to mean-field exchange interaction, gives rise to the additional exchange energy $\Delta E_\mathrm{exch}=UN_0N_1/L^3$ in the presence of fragmentation, when both $N_0$ and $N_1$ are macroscopically large. From the other hand, the condensate fragmentation allows the system to lower its kinetic energy at $Q_0<Q\leqslant N/2$ by the amount $\Delta E_\mathrm{kin}=-k_0^2(Q-Q_0)^2/4mQ_0$ [it is the difference between solid and dashed lines in the inset in Fig.~\ref{Fig4}(a)]. The total change of system energy $\Delta E=\Delta E_\mathrm{kin}+\Delta E_\mathrm{exch}$ when it passes from unfragmented ($N_0=N[1-(T/T_\mathrm{c})^{3/2}]$, $N_1=0$) to fragmented ($N_0=N[1-(T/T_\mathrm{c})^{3/2}]-(Q-Q_0)$, $N_1=Q-Q_0$) BEC is
\begin{equation}
\Delta E=-\frac{k_0^2(Q-Q_0)^2}{4mQ_0}+\frac{U}{L^3}(N-Q-Q_0)(Q-Q_0).\label{DeltaE1}
\end{equation}

To estimate the relative magnitudes of both terms in Eq.~(\ref{DeltaE1}), we take $Q=N/2$ where $\Delta E_\mathrm{kin}$ reaches the extreme value. Taking into account that $k_0=2\pi\hbar/L$, $U=4\pi a_\mathrm{s}\hbar^2/m$, where $a_\mathrm{s}$ is the s-wave scattering length, and using Eq.~(\ref{Q0}), we obtain in this case
\begin{align}
\Delta E|_{Q=N/2}&=\frac{\pi^2\hbar^2N}{2mL^2}\left[1-\left(\frac{T}{T_\mathrm{c}}\right)^{3/2}\right]^2\nonumber\\
&\times\left\{-\left(\frac{T_\mathrm{c}}{T}\right)^{3/2}+\frac{2Na_\mathrm{s}}{\pi L}\right\}.\label{DeltaE2}
\end{align}
The first term in the braces always dominate at low enough temperature, so the fragmentation becomes energetically favorable ($\Delta E|_{Q=N/2}<0$) even in the presence of interaction. With the typical parameters of atomic BECs \cite{Pethick,Dalfovo} $a_\mathrm{s}=2-100\,\mbox{nm}$, $L=10-50\,\mathrm{\mu m}$, $N=10^4-10^5$, we obtain the temperature when fragmentation occurs in the range $T/T_\mathrm{c}\sim0.02-1$. Thus we conclude that condensate fragmentation predicted by our analysis should not be necessarily suppressed by the interaction effects. At some combinations of realistic system parameters, the exchange interaction, which counteracts the fragmentation, can be even negligibly weak.

\section{Distributions of occupation numbers}\label{Sec5}

\subsection{Saddle-point calculation of distributions}

In this section, we calculate the probabilities $p(N_\mathbf{q})$ for $N_\mathbf{q}$ particles to occupy the $\mathbf{q}$th  single-particle state. Our main focus is the 0th and 1st states which can host BEC at $0\leqslant Q\leqslant N/2$, with the other states considered in the last subsection.

Our calculations are based on the saddle-point expression (\ref{p3}), so we find the probability distribution functions in the large-$N$ limit. The saddle point location $(\tilde{z}(N_\mathbf{q}),\tilde{\mathbf{v}}(N_\mathbf{q}))$ for this integral depends on $N_\mathbf{q}$ and is determined by the equations
\begin{align}
\frac{\partial\tilde{f}}{\partial z}&=-N+N_\mathbf{q}+\sum_{\mathbf{k}\neq\mathbf{q}}\tilde\nu_\mathbf{k}=0,\label{gder1}\\
\frac{\partial\tilde{f}}{\partial\mathbf{v}}&=\frac2R\left(-\mathbf{Q}+\mathbf{q}N_\mathbf{q}+\sum_{\mathbf{k}\neq\mathbf{q}}\mathbf{k}\tilde\nu_\mathbf{k}\right)=0,\label{gder2}
\end{align}
where
\begin{equation}
\tilde\nu_\mathbf{k}=\frac1{e^{\frac{\mathbf{k}^2-2\tilde{\mathbf{v}}(N_\mathbf{q})\cdot\mathbf{k}}R-\tilde{z}(N_\mathbf{q})}-1}.\label{tilde_nu}
\end{equation}
From the physical point of view, Eqs.~(\ref{gder1})--(\ref{gder2}) fix the total particle number $N-N_\mathbf{q}$ and total momentum $\mathbf{Q}-\mathbf{q}N_\mathbf{q}$ for a system of bosons having the dimensionless chemical potential $\tilde{z}$ in the frame moving with the dimensionless velocity $\tilde{\mathbf{v}}$. The occupation of the $\mathbf{q}$th level is frozen and treated as the external parameter $N_\mathbf{q}$, while mean occupations of other levels (\ref{tilde_nu}) are given by the Bose-Einstein distribution.

\begin{figure}[t]
\begin{center}
\includegraphics[width=0.9\columnwidth]{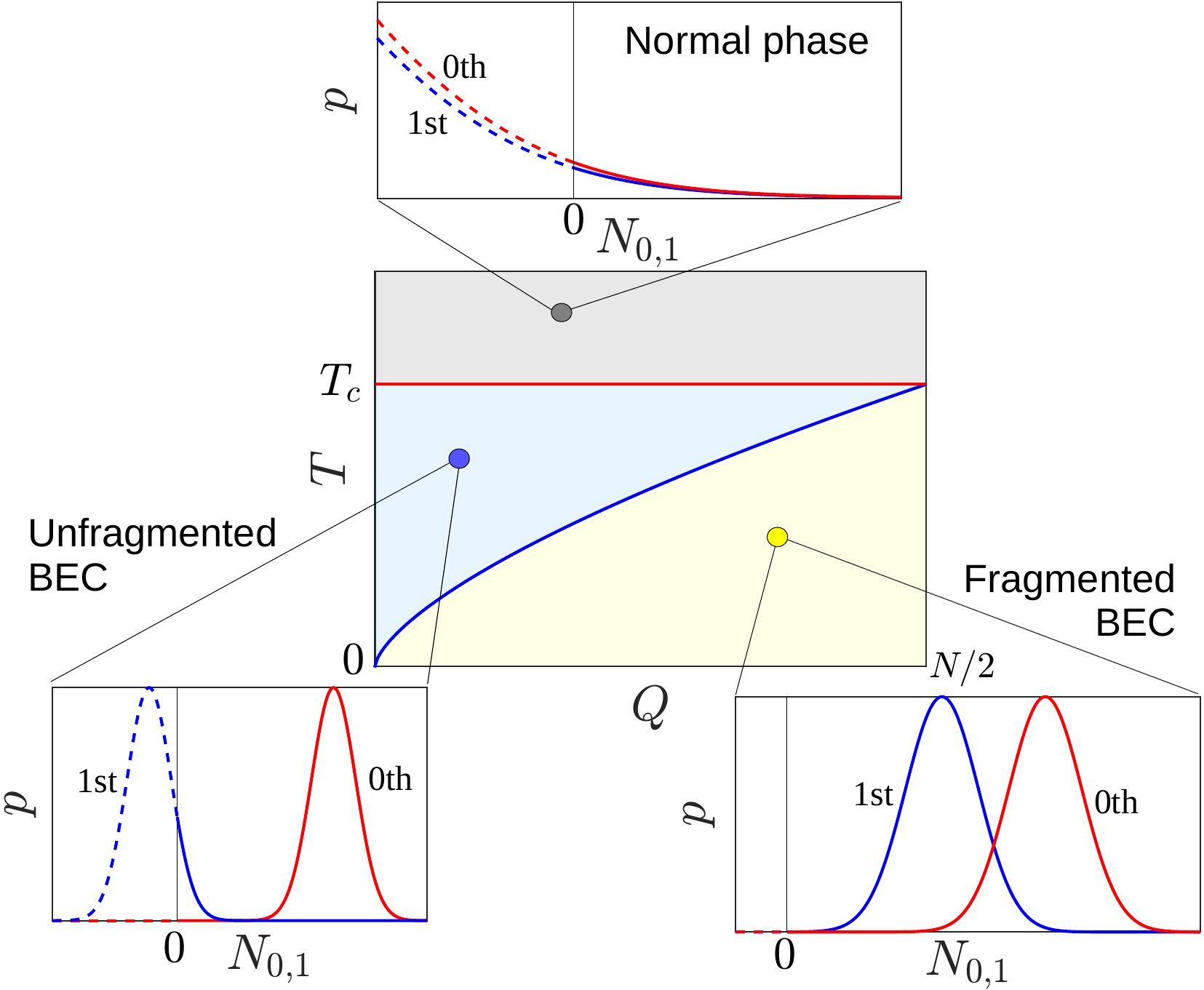}
\end{center}
\caption{\label{Fig5} Phase diagram in the range $0\leqslant Q\leqslant N/2$, where qualitative shapes of distribution functions $p(N_0)$ and $p(N_1)$ for occupations of the 0th and 1st single-particle levels are shown in different phases: normal phase $T>T_\mathrm{c}$, unfragmented BEC phase $T<T_\mathrm{c}$, $Q<Q_0$, and fragmented BEC phase $T<T_\mathrm{c}$, $Q>Q_0$. In all phases the saddle-point calculations yield approximately Gaussian shapes for $p(N_{0,1})$, but their parts at $N_{0,1}<0$, shown by dashed lines, are unphysical.}
\end{figure}

Further calculation of $p(N_\mathbf{q})$ is described in Appendix~\ref{Appendix_B} and consists in the following steps: 

i) for each $N_\mathbf{q}$ from Eqs.~(\ref{gder1})--(\ref{gder2}) we find the saddle point $(\tilde{z}(N_\mathbf{q}),\tilde{\mathbf{v}}(N_\mathbf{q}))$ whose vicinity provides the dominant contribution $p(N_\mathbf{q})\sim e^{\tilde{f}(N_\mathbf{q},\tilde{z}(N_\mathbf{q}),\tilde{\mathbf{v}}(N_\mathbf{q}))}$ to the integral (\ref{p3}); 

ii) among the occupation numbers $N_\mathbf{q}$ we find the value $\bar{N}_\mathbf{q}$ where $p(N_\mathbf{q})$ attains the maximum;

iii) we decompose the exponent in $p(N_\mathbf{q})$ around the maximum up to the term $(N_\mathbf{q}-\bar{N}_\mathbf{q})^2$ obtaining the Gaussian approximation for the distribution function; 

iv) in the case $\bar{N}_\mathbf{q}<0$, arising when condensate at the $\mathbf{q}$th level is absent, we approximate only the physically relevant tail $N_\mathbf{q}\geqslant0$ of the distribution by the exponentially decaying function.

Fig.~\ref{Fig5} depicts the general picture for distribution functions for 0th and 1st single-particle states. In the normal phase, both $p(N_0)$ and $p(N_1)$ attain their maxima at $\bar{N}_{0,1}<0$, so only their positive-$N_{0,1}$ exponential tails are physical. In the unfragmented BEC phase, $\bar{N}_0>0$ and $\bar{N}_1<0$, so $p(N_0)$ is Gaussian and $p(N_1)$ is exponential. In the fragmented BEC phase, both maxima are attained at $N_{0,1}>0$, thus both distribution functions $p(N_{0,1})$ are Gaussian.

\begin{figure}[t]
\begin{center}
\includegraphics[width=\columnwidth]{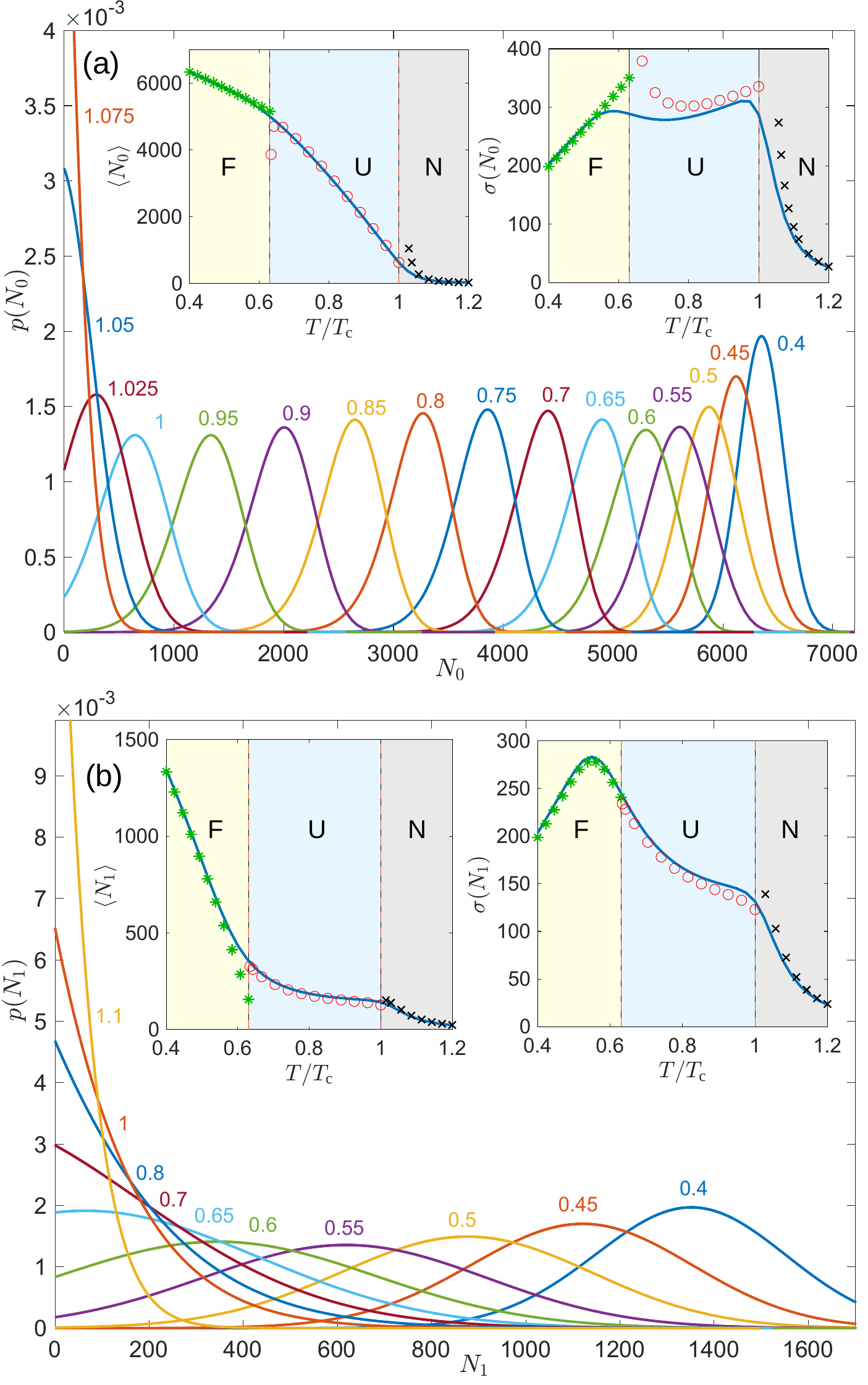}
\end{center}
\caption{\label{Fig6} Distribution functions $p(N_{0,1})$ for occupation numbers of (a) 0th and (b) 1st single-particle levels calculated at $N=10^4$, $Q=2500$ for different temperatures $T/T_\mathrm{c}$ indicated near corresponding curves. Insets show mean particle numbers $\langle N_{0,1}\rangle$ and their root-mean-square deviations $\sigma(N_{0,1})$ as functions of $T$, calculated numerically (solid lines) and approximated analytically (symbols, see text). Different approximations are used in the normal phase (N), unfragmented BEC phase (U), and fragmented BEC phase (F).}
\end{figure}

\begin{figure}[t]
\begin{center}
\includegraphics[width=\columnwidth]{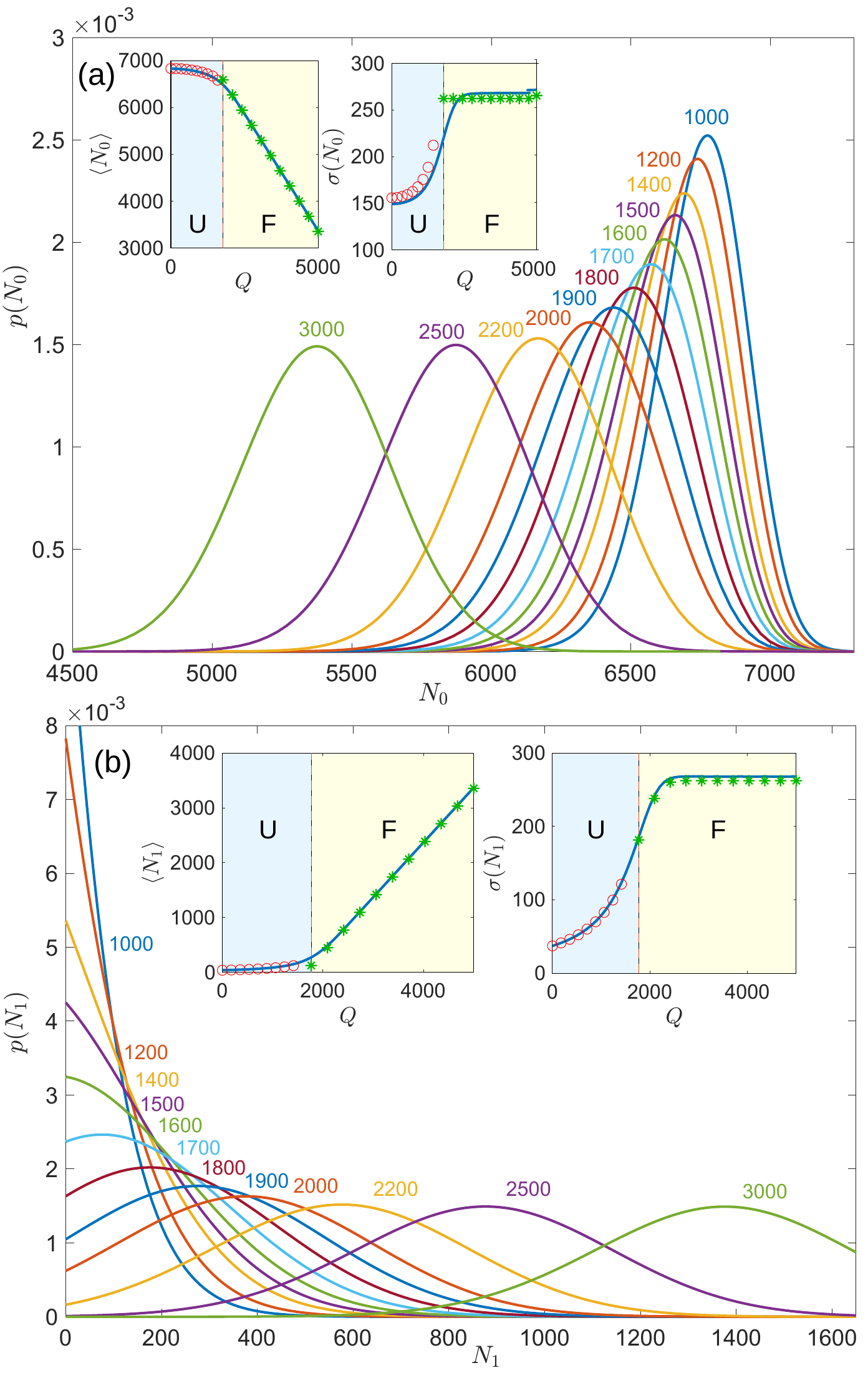}
\end{center}
\caption{\label{Fig7}  Distribution functions $p(N_\mathbf{q})$ for occupation numbers of (a) 0th and (b) 1st single-particle levels calculated at $N=10^4$, $T=0.5T_\mathrm{c}$ for different total momenta $Q$ indicated near corresponding curves. Insets show mean particle numbers $\langle N_{0,1}\rangle$ and root-mean-square deviations $\sigma(N_{0,1})$ as functions of $Q$, calculated numerically (solid lines) and approximated analytically (symbols, see text). Different approximations are used in the unfragmented (U) and fragmented (F) BEC phases.}
\end{figure}

Examples of numerically calculated distribution functions $p(N_0)$ and $p(N_1)$ are shown in Fig.~\ref{Fig6} at different $T$ and fixed $Q=0.25N$, and in Fig.~\ref{Fig7} at different $Q$ and fixed $T=0.5T_\mathrm{c}$. In the presence of a condensate at the $\mathbf{q}$th level, $p(N_\mathbf{q})$ has the Gaussian-like shape. Otherwise the distribution $p(N_\mathbf{q})$ is exponentially decaying at $N_\mathbf{q}\geqslant0$, as seen in Fig.~\ref{Fig6}(a) at $T>T_\mathrm{c}$, in Fig.~\ref{Fig6}(b) at $T>0.63T_\mathrm{c}$, and in Fig.~\ref{Fig7}(b) at $Q<1768$. 

Insets in Figs.~\ref{Fig6} and \ref{Fig7} show the mean occupation numbers $\langle N_\mathbf{q}\rangle$ of the 0th and 1st levels, and their root-mean-square deviations $\sigma(N_\mathbf{q})=\langle(N_\mathbf{q}-\langle N_\mathbf{q}\rangle)^2\rangle^{1/2}$, or particle-number fluctuations. The leading-order ($\sim N$) analytical results for $\langle N_{0,1}\rangle$ in different regimes (normal state as well as unfragmented and fragmented BEC) are given in Eqs.~(\ref{z_cond1})--(\ref{nu1_cond2}) by the quantities $\nu_{0,1}$. In the following subsections, we go beyond this simple analysis calculating $\langle N_{0,1}\rangle$ up to subleading ($\sim N^{2/3}$) corrections and estimating the distribution widths $\sigma(N_{0,1})$ in the leading order. To achieve this accuracy, sometimes we need to take into account subleading terms in the saddle-point integral (\ref{p3}) caused by pre-exponential factors, as described in details in Sec.~II of the Supplemental Material \cite{Supplemental}.

\subsection{Normal phase, $T>T_\mathrm{c}$}

Above the critical temperature, the condensate is absent on both 0th and 1st levels, which means that the distribution maxima are attained at negative $N_{0,1}$. Thus we expand $\tilde{f}(N_\mathbf{q},\tilde{z}(N_\mathbf{q}),\tilde{\mathbf{v}}(N_\mathbf{q}))$ around $N_{0,1}=0$ up to linear term to obtain
\begin{align}
p(N_0)&\sim\exp\{\tilde{z}(0)N_0\},\label{p0_norm}\\
p(N_1)&\sim\exp\left\{\left[\tilde{z}(0)+\frac{2\tilde{v}_x(0)-1}R\right]N_1\right\}.\label{p1_norm}
\end{align}
Here $\tilde{v}_x(0)=Q/N$, and the dimensionless chemical potential $\tilde{z}(0)$ is given by the equation
\begin{equation}
\frac1{\zeta(\frac32)}\left(\frac{T}{T_\mathrm{c}}\right)^{3/2}g_{3/2}(-\tilde{z}(0))=1.\label{zt_norm}
\end{equation}
Mean occupation numbers and their standard deviations calculated from Eqs.~(\ref{p0_norm})--(\ref{p1_norm}) are shown in Fig.~\ref{Fig6}(a) and Fig.~\ref{Fig6}(b) by black crosses.

\subsection{Unfragmented BEC, $T<T_\mathrm{c}$, $Q<Q_0$}

In this regime the condensate is present only at the 0th level. Its occupation probability $p(N_0)$ has the Gaussian shape
\begin{equation}
p(N_0)\sim\exp\left\{-\frac{\pi^2[\zeta(\frac32)]^{4/3}(N_0-\bar{N}_0)^2}{2N^{4/3}(T/T_\mathrm{c})^2J_1(Q/2Q_0,\alpha_0)}\right\}\label{p0_BEC1_distr}
\end{equation}
with the maximum at
\begin{equation}
\bar{N}_0=N\left[1-\left(\frac{T}{T_\mathrm{c}}\right)^{3/2}\right]-RI_1\left(\frac{Q}{2Q_0},\alpha_0\right)+o(N^{2/3}),\label{p0_BEC1_n0}
\end{equation}
where $\alpha_0$ is found from the equation
\begin{equation}
\alpha_0-\frac{J_2(Q/2Q_0,\alpha_0)}{2J_1^2(Q/2Q_0,\alpha_0)}=0.\label{p0_BEC1_max_cond}
\end{equation}
Here the integrals
\begin{align}
I_1(v_x,\alpha)&=\int\limits_0^\infty dt\:e^{\alpha t}\left[G(v_x,t)G^2(0,t)-1-\left(\frac\pi{t}\right)^{3/2}\right],\\
J_n(v_x,\alpha)&=\int\limits_0^\infty dt\:t^ne^{\alpha t}[G(v_x,t)G^2(0,t)-1]
\end{align}
are introduced, and $G(v,t)=\sum_{k=-\infty}^{\infty}e^{-k^2t+2kvt}$ is the Jacobi theta function. The analytical results for mean occupation (\ref{p0_BEC1_n0}) and the standard deviation provided by the Gaussian distribution (\ref{p0_BEC1_distr}) are shown in Fig.~\ref{Fig6}(a) and Fig.~\ref{Fig7}(a) by red circles.

At the 1st level, the condensate is absent, so we decompose the exponent $\tilde{f}(N_1,\tilde{z}(N_1),\tilde{\mathbf{v}}(N_1))$ around $N_1=0$, taking both the linear in $N_1$ and quadratic terms, because retaining only the former produces large numerical
errors. The resulting distribution is
\begin{equation}
p(N_1)\sim\exp\left\{-\frac{\pi[\zeta(\frac32)]^{2/3}(N_1-Q+Q_0)^2}{N^{5/3}(T/T_\mathrm{c})^{5/2}}\right\},
\end{equation}
and its mean and standard deviation calculated over $N_1\geqslant0$ are shown in Fig.~\ref{Fig6}(b) and Fig.~\ref{Fig7}(b) by red circles.

\subsection{Fragmented BEC, $T<T_\mathrm{c}$, $Q>Q_0$}

In this regime, both 0th and 1st levels host the condensate, so their occupation probabilities have the form of Gaussians
\begin{align}
p(N_0)&\sim\exp\left\{-\frac{\pi[\zeta(\frac32)]^{2/3}(N_0-\bar{N}_0)^2}{N^{5/3}(T/T_\mathrm{c})^{5/2}}\right\},\label{p0_BEC2_distr}\\
p(N_1)&\sim\exp\left\{-\frac{\pi[\zeta(\frac32)]^{2/3}(N_1-\bar{N}_1)^2}{N^{5/3}(T/T_\mathrm{c})^{5/2}}\right\},
\end{align}
centered around the most probable occupation numbers
\begin{align}
\bar{N}_0&=N\left[1-\left(\frac{T}{T_\mathrm{c}}\right)^{3/2}\right]-(Q-Q_0)+\frac{RI_2}2+o(N^{2/3}),\label{p0_BEC2_n0}\\
\bar{N}_1&=Q-Q_0+\frac{RI_2}2+o(N^{2/3}).\label{p1_BEC2_n0}
\end{align}
Here the integral
\begin{equation}
I_2=-\int\limits_0^\infty dt\left[G\left(\frac12,t\right)G^2(0,t)-2-\left(\frac\pi{t}\right)^{3/2}\right]\approx6.375\label{I3}
\end{equation}
was introduced.

Note that we can also calculate the joint distribution $p(N_0,N_1)$ of both occupation numbers in the fragmented BEC phase. Our estimates show that location of its maximum is close to those of marginal distributions (\ref{p0_BEC2_n0})--(\ref{p1_BEC2_n0}) except slight deviations of the order of $R$. Interestingly, this joint distribution is anisotropic:  in the large-$N$ limit its width along $N_0+N_1$, being of the order of $N^{2/3}$, is much smaller than the width along $N_0-N_1$ having the order $N^{5/6}$. In other words, the sum of occupation numbers $N_0+N_1$ fluctuates much weaker than their difference, or both $N_0$ and $N_1$ on their own.

\subsection{Other levels $\mathbf{q}\neq0,\mathbf{e}_x$}

Now consider the distribution functions $p(N_\mathbf{q})$ for occupation numbers on other levels $\mathbf{q}\neq0,\mathbf{e}_x$ which should not host the condensate at $0\leqslant Q\leqslant N/2$. Due to the absence of condensate, we need to expand $p(N_\mathbf{q})$ around $N_\mathbf{q}=0$, and with this condition the saddle-point equations (\ref{gder1})--(\ref{gder2}) reduce to those for partition function (\ref{fder3})--(\ref{fder4}), because exclusion of the momentum $\mathbf{q}$ from the sum over states provides only a small error $\mathcal{O}(1)$. Hence we can use the results (\ref{z_norm})--(\ref{v_norm}) and (\ref{z_cond1})--(\ref{nu1_cond2}) for the saddle-point parameters for partition function and substitute them to the general formula (\ref{p7}) for the distribution function to obtain: 
\begin{equation}
p(N_\mathbf{q})~\sim\exp\left\{\left[\tilde{z}(0)+\frac{-\mathbf{q}^2+2\tilde{v}_x(0)q_x}R\right]N_\mathbf{q}\right\}.\label{pq}
\end{equation}
In the normal phase $T>T_\mathrm{c}$, $\tilde{z}(0)$ is given by Eq.~(\ref{zt_norm}) and $\tilde{v}_x(0)=Q/N$. In the BEC phases $T<T_\mathrm{c}$, we take $\tilde{z}(0)\approx0$ and $\tilde{v}_x(0)=\min(Q/2Q_0,1/2)$.

Thus in both normal and BEC phases the distribution functions (\ref{pq}) for occupations of non-condensate levels $\mathbf{q}$ take the Gibbsian form $p(N_\mathbf{q})\sim\exp\{-[\beta\varepsilon_{\mathbf{q}-\tilde{\mathbf{v}}(0)}-\tilde{z}(0)]N_\mathbf{q}\}$ with suitable dimensionless chemical potential $\tilde{z}(0)$ and the reference frame boost $\tilde{\mathbf{v}}(0)$. Note that the levels whose momenta $\mathbf{q}$ are co-directional with $\tilde{\mathbf{v}}$ (and generally with the system total momentum $\mathbf{Q}$) become more populated to partly accommodate it, and those with the counter-directional momenta $\mathbf{q}$ become less populated.

\section{Effect of momentum fixation}\label{Sec6}

In this section, we will compare our analysis with the total momentum fixed to $\mathbf{Q}=0$ and conventional approach for ideal Bose gas in canonical ensemble without total momentum fixation. As shown in Sec. III of the Supplemental 
Material \cite{Supplemental}, the partition $Z$ and distribution $p(N_0)$ functions, calculated with the fixed total momentum $\mathbf{Q}=0$ using Eqs.~ (\ref{Z_NQ3}) and (\ref{p3}), differ from those calculated in canonical ensemble only by extra integration over $\mathbf{v}=0$ around the saddle point $(z_0,\mathbf{v}=0)$ and $(\tilde{z}(N_\mathbf{q}),\tilde{\mathbf{v}}=0)$, respectively. The resulting difference in $Z$ and $p(N_0)$ turns our to be relatively small when $T/T_\mathrm{c}\gg N^{-2/3}$, i.e. in the thermodynamic limit $N\rightarrow\infty$ taken at fixed temperature. 

\begin{figure}[t]
\begin{center}
\includegraphics[width=\columnwidth]{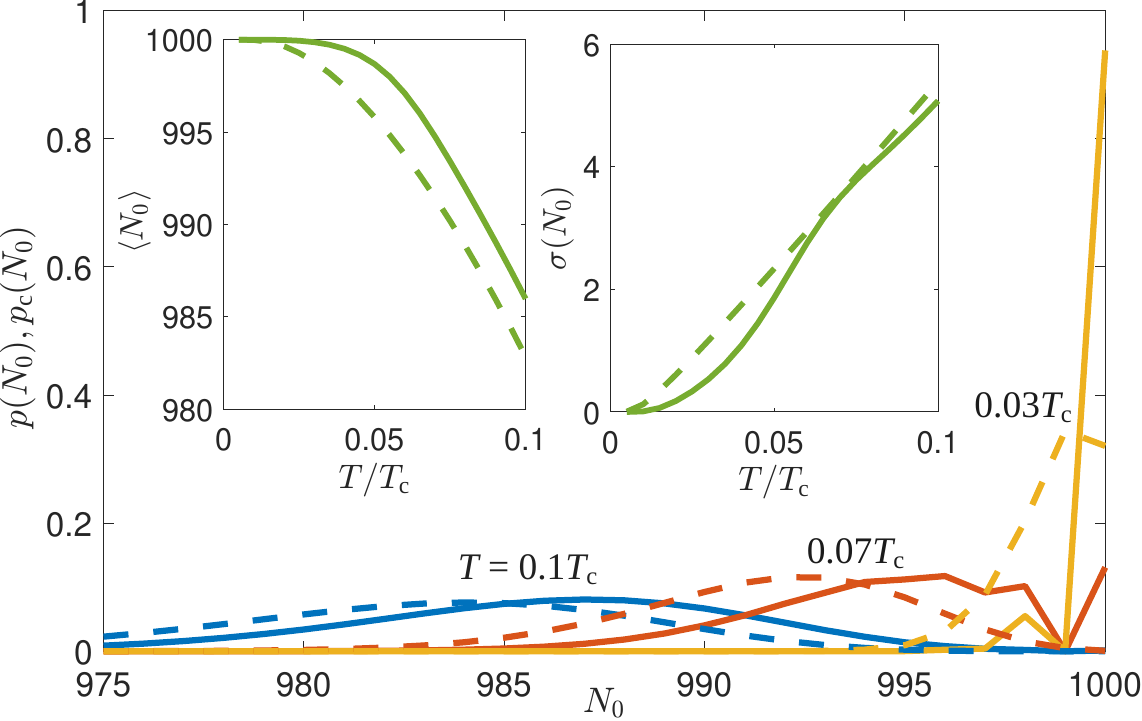}
\end{center}
\caption{\label{Fig8} Distribution functions $p(N_0)$ (solid lines) and $p_\mathrm{c}(N_0)$ (dashed lines) for occupation number $N_0$ of the 0th condensate level calculated at $N=1000$ for different temperatures showed near the curves. Insets show mean particle number $\langle N_0\rangle$ and root-mean-square deviations $\sigma(N_0)$ as functions of $T$. Solid lines correspond to calculations at fixed total momentum $Q=0$, and dashed lines correspond to canonical ensemble without momentum fixation.}
\end{figure}

At lower temperatures, when $T/T_\mathrm{c}\lesssim N^{-2/3}$, fixation of zero total momentum $\mathbf{Q}=0$ significantly changes both free energy and distribution function $p(N_0)$. In particular, $p(N_0)$ becomes much more narrow than its counterpart $p_\mathrm{c}(N_0)$ calculated in canonical ensemble. This is shown in Fig.~\ref{Fig8}, where the distribution functions are compared at $N=1000$ and at several temperatures approaching the low-temperature condition $T/T_\mathrm{c}\sim N^{-2/3}=0.01$. As seen,  at progressively lower $T$ the total-momentum fixed distribution $p(N_0)$ (solid lines) becomes more and more narrow in comparison with the canonical ensemble result $p_\mathrm{c}(N_0)$ (dashed lines).

We should note that the saddle-point method itself, which is used in our calculations, becomes poorly applicable at $T/T_\mathrm{c}\lesssim N^{-2/3}$, because the integrands in Eqs.~(\ref{Z_NQ3})--(\ref{p3}) become relatively wide and essentially non-Gaussian near the saddle points. Besides, we cannot treat the parameter $R$ [see Eq.~(\ref{R})] as large in this case. At extremely low temperatures $T/T_\mathrm{c}\ll N^{-2/3}$, when $R\ll1$, it is better to calculate the distribution functions $p(N_0)$ and $p_\mathrm{c}(N_0)$ by direct summation (\ref{p1}) over 7 single-particle states $\mathbf{q}=0,\pm\mathbf{e}_x,\pm\mathbf{e}_y,\pm\mathbf{e}_z$, which are closest to the origin. In this approximation, we obtain the average occupation $\langle N_0\rangle=N-6e^{-2/R}$ and and its root-mean-square deviation $\sigma(N_0)=2\sqrt3e^{-1/R}$ with the fixed total momentum $Q=0$. To compare, in canonical ensemble we obtain $\langle N_0\rangle_\mathrm{c}=N-6e^{-1/R}$ and $\sigma_\mathrm{c}(N_0)=\sqrt6e^{-1/2R}$. Indeed, in the insets of Fig.~\ref{Fig8} we see that at $T\rightarrow0$ the mean occupation of the 0th level in canonical ensemble is lower, $\langle N_0\rangle_\mathrm{c}<\langle N_0\rangle$, and its width is larger, $\sigma_\mathrm{c}(N_0)>\sigma(N_0)$, because of less restricted distribution of particles among the single-particles levels than in the presence of momentum fixation.

The related question is how the total momentum $\mathbf{Q}$ fluctuates in canonical ensemble when it is not fixed. Taking into account that partition function in canonical ensemble reads
\begin{equation}
Z_{\mathrm{c}}=\sum_{\{n_\mathbf{k}\}}e^{-\beta\sum\limits_\mathbf{k}n_\mathbf{k}\varepsilon_\mathbf{k}}\delta_{N,\sum\limits_\mathbf{k}n_\mathbf{k}}\label{Z_N}
\end{equation}
and comparing this formula with Eq.~(\ref{Z_NQ1}), we see that probability of the unrestricted system to have the total momentum $\mathbf{Q}$ is
\begin{equation}
p(\mathbf{Q})=\frac{Z}{Z_{\mathrm{c}}}=e^{-\frac{F(N,\mathbf{Q})-F_\mathrm{c}(N)}T}.\label{Q_distr1}
\end{equation}
In the normal state $T>T_\mathrm{c}$, the free energy $F(N,\mathbf{Q})$ of the fixed-momentum system given by Eq.~(\ref{F_norm}) is higher than those in canonical ensemble $F_\mathrm{c}(N)$ [equal to the first term in Eq.~(\ref{F_norm})] by $k_0^2\mathbf{Q}^2/2mN$, so the distribution (\ref{Q_distr1}) has the ordinary Maxwellian form,
\begin{equation}
p(\mathbf{Q})\sim e^{-\mathbf{Q}^2/NR}=e^{-\mathbf{P}^2/2mNT},
\end{equation}
for the system of $N$ particles taken as a whole. In the unfragmented BEC state $T<T_\mathrm{c}$, $|Q_{x,y,z}|<Q_0$, Eq.~(\ref{F_BEC1}) yields the similar difference of free energies $k_0^2\mathbf{Q}^2/2mN'$, but with the smaller number $N'=N(T/T_\mathrm{c})^{3/2}$ in the denominator, equal to particle number in the thermal cloud. The resulting distribution in this case,
\begin{equation}
p(\mathbf{Q})\sim e^{-(\mathbf{Q}^2/NR)(T_\mathrm{c}/T)^{3/2}}=e^{-(\mathbf{P}^2/2mNT)(T_\mathrm{c}/T)^{3/2}},\label{pQ_BEC1}
\end{equation}
is more narrow than in the normal state, because in the Bose-condensed system the Galilean invariance is broken, and the center-of-mass motion is no longer decoupled from the relative motion. Eq.~(\ref{pQ_BEC1}) is formally applicable only at low enough total momenta $|\mathbf{Q}|\lesssim Q_0$. At larger momenta we enter the fragmented BEC regime $T<T_\mathrm{c}$, $|Q_i|>Q_0$, where the difference of the momentum-fixed (\ref{F_BEC2}) and unrestricted free energies switch from quadratic to linear dependence on $|\mathbf{Q}|$ [see Fig.~\ref{Fig4}(a)], so the distribution $p(\mathbf{Q})$ switches from Gaussian to exponential. However, if $e^{-(Q_0^2/NR)(T_\mathrm{c}/T)^{3/2}}\ll1$, the exponential high-momentum tail can be neglected. Taking into account Eq.~(\ref{R}), this happens at $T/T_\mathrm{c}\gtrsim N^{-2/3}$, when temperature is not very small and the saddle-point method is still applicable. Thus we have found that at moderate temperatures $N^{-2/3}\lesssim T/T_\mathrm{c}<1$ the distribution of fluctuating total momentum $\mathbf{P}$ in canonical ensemble is Gaussian, although involves only the thermal cloud.

\section{Conclusions}\label{Sec7}

We have considered Bose condensation of ideal three-dimensional gas with fixed particle number $N$ and fixed total momentum $\mathbf{P}$. In principle, partition function and all other thermodynamic properties of the system can be found in this setting using the recurrence relations (\ref{Z_rec})--(\ref{p_rec}), but their application is limited to $N$ not exceeding several hundreds. To uncover universal features arising in the large-$N$ limit, we integrate over complex chemical potential $z$ and complex rapidity $\mathbf{v}$ to fix both $N$ and $\mathbf{P}$. In the large-$N$ limit, the dominating contribution to the integral is provided by vicinity of the saddle point $(z_0,\mathbf{v}_0)$. Interestingly, the saddle-point conditions look like transition to the grand canonical ensemble with definite chemical potential $\mu=Tz_0+k_0^2\mathbf{v}_0^2/2m$ in the reference frame moving with the velocity $k_0\mathbf{v}_0/m$, where $k_0=2\pi\hbar/L$ is the quantization unit of momentum in the system of size $L$.

Studying the system phase diagram with respect to $T$ and dimensionless total momentum $Q=P_x/k_0$ along the $x$ axis, we identify three different phases: (1) normal phase at $T>T_\mathrm{c}$, where $T_\mathrm{c}$ is the conventional critical temperature of BEC (\ref{T_c}); (2) unfragmented BEC phase at $T<T_\mathrm{c}$, $0\leqslant Q<Q_0$ with the threshold momentum $Q_0$ (\ref{Q0}), where only the $\mathbf{k}=0$ level hosts the condensate; (3) fragmented BEC phase at $T<T_\mathrm{c}$, $Q_0<Q\leqslant N/2$, where the condensate, i.e. mean macroscopic occupation, is present at the $\mathbf{k}=0$ and $\mathbf{k}=\mathbf{e}_x$ levels simultaneously. Fragmentation of the condensate can be explained using energetic arguments: at large enough total momentum, macroscopic occupation of the moving $\mathbf{k}=\mathbf{e}_x$ state becomes more energetically favorable for the particles than staying in the non-condensed thermal cloud. Although usually the fragmentation is suppressed by exchange energy \cite{Nozieres}, our estimates show that this its effect can be weak at realistic parameters of atomic BECs, especially at low temperatures. Beyond the range $0\leqslant Q\leqslant N/2$, the sequence of unfragmented and fragmented phases is repeated with the period $\Delta Q=N$ and symmetry around $Q=0$, although with varying specific numbers of condensate levels. At arbitrary three-dimensional momenta $\mathbf{P}$, we can expect up to 8 single-particle levels hosting the condensate.

Saddle point method allowed us to derive analytical expressions for partition function $Z$ and probability distributions $p(N_\mathbf{k})$ for occupation numbers of the $\mathbf{k}$th single-particle states in all aforementioned phases. These expressions become asymptotically exact when the parameter $R\propto N^{2/3}T/T_\mathrm{c}$ tends to infinity, i.e. in the thermodynamic limit $N,L\rightarrow\infty$ at $N/L^3=\mathrm{const}$ and $T=\mathrm{const}$ (although numerical saddle-point integration provide accurate results already at $R>2$). The distribution $p(N_\mathbf{k})$ has approximately exponential shape for noncondensate levels and Gaussian shape for those levels $\mathbf{k}$ which host the condensate $\langle N_\mathbf{k}\rangle\sim N$. In the latter case, we deduce both leading ($\sim N$) and subleading ($\sim N^{2/3}$) terms in analytical expressions for the mean value $\langle N_\mathbf{k}\rangle$ to achieve better accuracy. Fluctuations of the number of Bose-condensed particles at each condensate level are given by the standard deviation of the order of $N^{2/3}$ in the unfragmented phase and $N^{5/6}$ in the fragmented phase. In the latter case, as we verified by additional analysis of the joint distribution $p(N_0,N_1)$, the total number of condensate particles $N_0+N_1$ fluctuate weaker (with the standard deviation $\sim N^{2/3}$, as for unfragmented BEC) than each of the numbers $N_{0,1}$ separately. Note that distribution functions can be related as $p(N_\mathbf{q})\propto\exp(-\beta L(\phi_\mathbf{q}))$ to Landau functionals $L(\phi_\mathbf{q})$ of the corresponding BEC order parameters having amplitudes $|\phi_\mathbf{q}|=\sqrt{N_\mathbf{q}}$ \cite{Sinner}, so the Gaussian and exponential shapes of $p(N_\mathbf{q})$ correspond to minima of $L(\phi_\mathbf{q})$ at, respectively, $|\phi_\mathbf{q}|\neq0$ and $|\phi_\mathbf{q}|=0$.

We also compared the thermodynamic (free energy) and statistical (occupation number distribution functions) properties of the system with the fixed total momentum $\mathbf{P}=0$ and without momentum fixation, in ordinary fixed-$N$ canonical ensemble. These two cases are shown to essentially differ only at low enough temperatures $T/T_\mathrm{c}\lesssim N^{-2/3}$ when only few lowest-energy single-particle states are significantly occupied, and discreteness of particle momenta in a finite system cannot be neglected. At the same time, at such low temperatures the saddle point method becomes poorly applicable because of essentially non-Gaussian integration. Thus we show that standard canonical ensemble treatment of BEC provides fairly good approximation for the system at rest, at strictly zero total momentum if the temperature is much higher than quantization scale of kinetic energy. However, at $\mathbf{P}\neq0$ it is not applicable any more, and the condensate fragmentation effects witnessing breaking of Galilean invariance in Bose-condensed system at $T<T_\mathrm{c}$ arise. Another signature of this invariance breaking is suppression of the center-of-mass momentum fluctuations in canonical ensemble in the presence of BEC at $T<T_\mathrm{c}$.

Our analysis extends the statistical theory of BEC in canonical ensemble \cite{Kocharovsky2006,Svidzinsky,Cockburn,Kocharovsky2010,Borrmann,Wang,Weiss,Holthaus,Wang_Ma,Idziaszek,Kocharovsky_PRL} by including the additional conserving quantity $\mathbf{P}$ besides $N$. On the other hand, it can be considered as extension of quantum theory of finite-$N$ Bose systems with nonzero angular momentum \cite{Yannouleas,Cooper,Alon} to the finite-temperature and large-$N$ limit because of similarity between angular momentum in annular geometry and total momentum at periodic boundary conditions. We consider noninteracting bosons, so our calculations cannot properly describe superfluidity and formation of metastable currents \cite{Bloch} as well as obstacles to the condensate fragmentation due to exchange effects \cite{Nozieres}, arising in the presence of repulsive interaction or nonlinearities which can reach even single-photon level at room temperature \cite{Zasedatelev}. Nevertheless, the predicted behaviour of moving Bose-condensed system can be checked in experiments with cold atomic gases where Feshbach resonances allow to suppress interaction \cite{Weber}. Possible future development of our approach can include taking into account interactions using the mean-field Bogoliubov theory, considering microcanonical ensemble with fixation of the total energy instead of temperature, and explicit consideration of rotational geometries of trapped gases.

\section*{Acknowledgments}
The work was done as a part of research Project No. FFUU-2024-0003 of the Institute for Spectroscopy of the Russian Academy of Sciences. The work on numerical calculations was supported by the Program of Basic Research of the
Higher School of Economics.

\appendix

\section{Large-$N$ limit of the logarithmic\\
sum and its derivatives}\label{Appendix_A}

Conventional treatment of the momentum sums like those appearing in Eqs.~(\ref{fder1})--(\ref{fder2}) and (\ref{gder1})--(\ref{gder2}) relies on switching from summation to integration over $\mathbf{k}$. The terms corresponding to the single-particle levels $\mathbf{k}$ which are expected to be macroscopically occupied should be treated separately. For the sum of logarithms in Eqs.~(\ref{f1})--(\ref{g1}), there is no need for such separation because any single term can provide, at most, only subextensive $\mathcal{O}(\log N)$ contribution. Switching from summation to integration and integrating by parts over $|\mathbf{k}|$, we obtain
\begin{align}
&-\sum_\mathbf{k}\log\left(1-e^{\frac{-\mathbf{k}^2+2\mathbf{v}\cdot\mathbf{k}}R+z}\right)\nonumber\\
&=\frac{N}{\zeta\left(\frac32\right)}\left(\frac{T}{T_\mathrm{c}}\right)^{3/2}g_{5/2}\left(-\frac{\mathbf{v}^2}R-z\right)+\mathcal{O}(\log N).\label{log_sum}
\end{align}
Here the Bose-Einstein integrals \cite{Robinson}
\begin{equation}
g_d(x)=\frac1{\Gamma(d)}\int\limits_0^\infty dx\:\frac{x^{d-1}}{e^{x+z}-1}=\sum_{s=1}^\infty\frac{e^{-sx}}{s^d}\label{BE_int}
\end{equation}
were introduced, which have the properties
\begin{align}
g_d'(x)&=-g_{d-1}(x),\\
g_d(0)&=\zeta(d)\mbox{ when }d>0,
\end{align}
and $\zeta(x)$ is the Riemann zeta function. 

The saddle-point conditions (\ref{fder1})--(\ref{fder2}) need more accurate treatment because the 0th ($\mathbf{q}=0$) and 1st ($\mathbf{q}=\mathbf{e}_x$) levels, which can potentially host the condensate, can provide large, linear in $N$, contributions to the derivatives of the momentum sum (\ref{log_sum}). Let us denote the set of such \emph{condensate} levels by $C$ and consider the sum over remaining \emph{noncondensate} levels $\mathbf{k}\notin C$. As shown in more details in Sec.~I of the Supplemental Material \cite{Supplemental}, we can approximate these derivatives as
\begin{align}
-\frac\partial{\partial z}\sum_{\mathbf{k}\notin C}&\log\left(1-e^{\frac{-\mathbf{k}^2+2\mathbf{v}\cdot\mathbf{k}}R+z}\right)\nonumber\\
&=\frac{N}{\zeta(\frac32)}\left(\frac{T}{T_\mathrm{c}}\right)^{3/2}g_{3/2}(-z)+\mathcal{O}(N^{2/3}),\label{log_sum_der1}\\
-\frac\partial{\partial v_x}\sum_{\mathbf{k}\notin C}&\log\left(1-e^{\frac{-\mathbf{k}^2+2\mathbf{v}\cdot\mathbf{k}}R+z}\right)\nonumber\\
&=\frac2R\left\{\frac{v_xN}{\zeta(\frac32)}\left(\frac{T}{T_\mathrm{c}}\right)^{3/2}g_{3/2}(-z)+\mathcal{O}(N^{2/3})\right\}.\label{log_sum_der2}
\end{align}
From the physical point of view, Eq.~(\ref{log_sum_der1}) provides the population of all noncondensate levels at the saddle point. The first term in the right hand side provides the leading ($\sim N$) order of a particle number in the thermal cloud in the thermodynamic limit, and the $\mathcal{O}(N^{2/3})$ term is responsible for the finite-size correction to it. Similarly, Eq.~(\ref{log_sum_der2}) presents the total momentum (times $2/R$) of all noncondensate states which comprise the thermal cloud and move with the average rapidity $v_x$. Using Eqs.~(\ref{log_sum_der1}) and (\ref{log_sum_der2}) with $C=\{0,\mathbf{e}_x\}$ in Eq.~(\ref{f1}), we obtain the saddle-point conditions (\ref{fder3}) and (\ref{fder4}), respectively.

\section{Calculation of distribution functions}\label{Appendix_B}

Particle number distribution function $p(N_\mathbf{q})$ for $\mathbf{q}=0$ or $\mathbf{q}=\mathbf{e}_x$ is given by the four-dimensional integral (\ref{p3}), which is dominated in the large-$N$ limit by a Gaussian integral in vicinity of the saddle point $(\tilde{z}(N_\mathbf{q}),\tilde{\mathbf{v}}(N_\mathbf{q}))$, whose location depends on $N_\mathbf{q}$:
\begin{equation}
p(N_\mathbf{q})\sim \frac1{\sqrt{H}}e^{\tilde{f}(N_\mathbf{q},\tilde{z}(N_\mathbf{q}),\tilde{\mathbf{v}}(N_\mathbf{q}))}.\label{p4}
\end{equation}
Here we can write the function (\ref{g1})
\begin{equation}
\tilde{f}(N_\mathbf{q},z,\mathbf{v})=h(z,\mathbf{v})+zN_\mathbf{q}+\frac{(-\mathbf{q}^2+2\mathbf{v}\cdot\mathbf{q})N_\mathbf{q}}R,\label{g2}
\end{equation}
in terms of
\begin{equation}
h(z,\mathbf{v})=-zN-\frac{2\mathbf{v}\cdot\mathbf{Q}}R-\sum_{\mathbf{k}\neq\mathbf{q}}\log\left(1-e^{\frac{-\mathbf{k}^2+2\mathbf{v}\cdot\mathbf{k}}R+z}\right)\label{h}.
\end{equation}
The Hessian determinant
\begin{equation}
H=\left.\left\{\frac{\partial^2h}{\partial z^2}\frac{\partial^2h}{\partial v_x^2}-\left(\frac{\partial^2h}{\partial z\partial v_x}\right)^2\right\}\right|\strut_{z=\tilde{z}(N_\mathbf{q})\atop\mathbf{v}=\tilde{\mathbf{v}}}\label{Hessian}
\end{equation}
of both $\tilde{f}$ and $h$, calculated at the saddle point, provides subleading contribution to (\ref{p4}) with respect to the leading-order exponential term. We retain it to obtain refined expressions for locations of the distribution maxima.

The saddle point location $(\tilde{z}(N_\mathbf{q}),\tilde{\mathbf{v}}(N_\mathbf{q}))$ is defined by stationarity conditions (\ref{gder1})--(\ref{gder2}), which can be written as
\begin{equation}
\frac{\partial\tilde{f}}{\partial z}=\frac{\partial h}{\partial z}+N_\mathbf{q}=0,\quad
\frac{\partial\tilde{f}}{\partial\mathbf{v}}=\frac{\partial h}{\partial\mathbf{v}}+\frac{2\mathbf{q}N_\mathbf{q}}R=0.\label{gder4}
\end{equation}
The maximum of distribution function (\ref{p4}) is \emph{formally} achieved at some particle number $N_\mathbf{q}=\bar{N}_\mathbf{q}$, where
\begin{align}
&\left.\frac{d}{dN_\mathbf{q}}\left(\tilde{f}-\frac12\log H\right)\right|\strut_{z=\tilde{z}(N_\mathbf{q})\atop\mathbf{v}=\tilde{\mathbf{v}}(N_\mathbf{q})}\nonumber\\
&\quad=\frac{-\mathbf{q}^2+2\tilde{\mathbf{v}}(N_\mathbf{q})\cdot\mathbf{q}}R+\tilde{z}(N_\mathbf{q})-\frac12\frac{d\log H}{dN_\mathbf{q}}=0.\label{max_cond}
\end{align}
Here we have used the stationarity conditions (\ref{gder4}). The \emph{actual} location of the maximum coincides with $\bar{N}_\mathbf{q}$ if $\bar{N}_\mathbf{q}>0$, so in this case we can perform Gaussian expansion of $p(N_\mathbf{q})$ around $\bar{N}_\mathbf{q}$. Otherwise, if $\bar{N}_\mathbf{q}<0$, only a right tail of this Gaussian distribution will be present in the physical region $N_\mathbf{q}\geqslant0$, so in the large-$N$ limit we can retain only its exponential asymptotic.

Thus, in the case $\bar{N}_\mathbf{q}>0$ we expand the distribution (\ref{p4}) around the maximum $N_\mathbf{q}=\bar{N}_\mathbf{q}$ in the Gaussian form
\begin{equation}
p(N_\mathbf{q})\sim \exp\left\{\frac12\left.\frac{d^2\tilde{f}}{dN_\mathbf{q}^2}\right|\strut_{z=\tilde{z}(\bar{N}_\mathbf{q})\atop\mathbf{v}=\tilde{\mathbf{v}}(\bar{N}_\mathbf{q})}(N_\mathbf{q}-\bar{N}_\mathbf{q})^2\right\}.\label{p5}
\end{equation}
To calculate the second derivative $d^2\tilde{f}/dN_\mathbf{q}^2$ at the saddle point we take into account Eqs.~(\ref{g2})--(\ref{h}) together with the stationarity conditions (\ref{gder4}) and their derivatives with respect to $N_\mathbf{q}$:
\begin{equation}
\left.\frac{d^2\tilde{f}}{dN_\mathbf{q}^2}\right|\strut_{z=\tilde{z}(\bar{N}_\mathbf{q})\atop\mathbf{v}=\tilde{\mathbf{v}}(\bar{N}_\mathbf{q})}=\frac{d\tilde{z}}{dN_\mathbf{q}}+\frac{2\mathbf{q}}R\cdot\frac{d\tilde{\mathbf{v}}}{dN_\mathbf{q}}.\label{gder6}
\end{equation}
From the other hand, differentiating Eq.~(\ref{gder4}) by $N_\mathbf{q}$ and taking into account that the mixed second derivatives $\partial^2h/\partial z\partial v_{y,z}$ and $\partial^2h/\partial v_i\partial v_j$ (at $i\neq j$) vanish at the saddle point since $\tilde{v}_y=\tilde{v}_z=0$, we obtain
\begin{align}
\frac{d\tilde{z}}{dN_\mathbf{q}}&=\frac1H\left(\frac{2q_x}R\frac{\partial^2h}{\partial z\partial v_x}-\frac{\partial^2h}{\partial v_x^2}\right),\label{spder1}\\
\frac{d\tilde{\mathbf{v}}}{dN_\mathbf{q}}&=\frac{\mathbf{e}_x}{H}\left(-\frac{2q_x}R\frac{\partial^2h}{\partial z^2}+\frac{\partial^2h}{\partial z\partial v_x}\right).\label{spder2}
\end{align}
Substituting these expressions to Eqs.~(\ref{gder6}) and (\ref{p5}), we get
\begin{align}
p(N_\mathbf{q})
\sim\exp&\left\{-\left(\frac{2q_x^2}{R^2}\frac{\partial^2h}{\partial z^2}-\frac{2q_x}R\frac{\partial^2h}{\partial z\partial v_x}+\frac12\frac{\partial^2h}{\partial v_x^2}\right)\vphantom{\frac{(N_\mathbf{q}-\bar{N}_\mathbf{q})^2}H}\right.\nonumber\\
&\;\;\left.\times\frac{(N_\mathbf{q}-\bar{N}_\mathbf{q})^2}H\right\}.\label{p6}
\end{align}
In the formulas (\ref{spder1})--(\ref{p6}), all derivatives of $h$ should be evaluated at the saddle point $(\tilde{z}(\bar{N}_\mathbf{q}),\tilde{\mathbf{v}}(\bar{N}_\mathbf{q}))$.

In the case $N_\mathbf{q}<0$ we expand $\tilde{f}$ in Eq.~(\ref{p4}) up to the linear term around $N_\mathbf{q}=0$ using the maximum condition (\ref{max_cond}) to obtain the exponential distribution:
\begin{equation}
p(N_\mathbf{q})
\sim\exp\left\{\left[\tilde{z}(0)+\frac{-\mathbf{q}^2+2\tilde{\mathbf{v}}(0)\cdot\mathbf{q}}R\right]N_\mathbf{q}\right\}.\label{p7}
\end{equation}

\bibliography{References}

\begin{thebibliography}{42}
\bibitem [{Sup()}]{Supplemental}%
  \BibitemOpen
  \href@noop {} {\bibinfo {title} {See {S}upplemental {M}aterial at [url will
  be inserted by publisher] for calculation details.}}\BibitemShut {Stop}%
\end{thebibliography}

\end{document}


\renewcommand{\theequation}{S.\arabic{equation}}

\title{Supplemental Material for ``Bose condensation in canonical ensemble\\with fixed total momentum''}
\author{Andrey S. Plyashechnik}%
\email{asplyashechnik@vniia.ru}
\affiliation{N.~L. Dukhov Research Institute of Automatics (VNIIA), 127055 Moscow, Russia}%
\author{Alexey A. Sokolik}%
\email{asokolik@hse.ru}%
\affiliation{Institute for Spectroscopy RAS, 142190 Troitsk, Moscow, Russia}%
\affiliation{National Research University Higher School of Economics, 109028 Moscow, Russia}%
\author{Yurii E. Lozovik}%
\affiliation{Institute for Spectroscopy RAS, 142190 Troitsk, Moscow, Russia}%
\affiliation{National Research University Higher School of Economics, 109028 Moscow, Russia}%
\affiliation{N.~L. Dukhov Research Institute of Automatics (VNIIA), 127055 Moscow, Russia}%

\maketitle

\section{Large-$N$ limits of logarithmic sum and its derivatives}

To simplify treatment of the momentum sums appearing in functions $f$ and $\tilde{f}$, we decompose the logarithms into Taylor series and use the Jacobi theta functions
\begin{equation}
G(v,t)=\sum_{k=-\infty}^{\infty}e^{-k^2t+2kvt}\label{Jacobi}
\end{equation}
to obtain
\begin{equation}
-\sum_\mathbf{k}\log\left(1-e^{\frac{-\mathbf{k}^2+2\mathbf{v}\cdot\mathbf{k}}R+z}\right)=\sum_{s=1}^\infty\frac{e^{sz}}s\prod_{i=x,y,z}G\left(v_i,\frac{s}R\right).\label{log_sum1}
\end{equation}
For the following analysis, it is useful to notice the asymptotic properties of the Jacobi functions at small and large $t$:
\begin{equation}
G(v,t)=\left\{\begin{array}{ll}e^{v^2t}\sqrt{\pi/t}[1+\mathcal{O}(e^{-\pi^2/t})],&t\ll1,\\e^{-k^2t+2kvt}+e^{-(k+1)^2t+2(k+1)vt}+o(e^{v^2t-t}),&t\gg1,\;k\leqslant v\leqslant k+1.\end{array}\right.\label{Jacobi_asympt}
\end{equation}
Subtracting the small-$t$ (or $s\lesssim R$) asymptotic $(\pi R/s)^{3/2}e^{\frac{s\mathbf{v}^2}R}$ of the product of Jacobi functions in Eq.~(\ref{log_sum1}) allows to calculate analytically the dominating contribution to the sum coming from $s\lesssim R$, that remains in the thermodynamic limit. Using the definition of parameter $R=N^{2/3}T/\pi[\zeta(\frac32)]^{2/3}T_\mathrm{c}$, we obtain
\begin{equation}
-\sum_\mathbf{k}\log\left(1-e^{\frac{-\mathbf{k}^2+2\mathbf{v}\cdot\mathbf{k}}R+z}\right)=\frac{N}{\zeta(\frac32)}\left(\frac{T}{T_\mathrm{c}}\right)^{3/2}g_{5/2}\left(-z-\frac{\mathbf{v}^2}R\right)+\sum_{s=1}^\infty\frac{e^{sz}}s\left[\prod_{i=x,y,z}G\left(v_i,\frac{s}R\right)-\left(\frac{\pi R}s\right)^{3/2}e^{\frac{s\mathbf{v}^2}R}\right],\label{log_sum2}
\end{equation}
where the Bose-Einstein integrals
\begin{equation}
g_d(x)=\frac1{\Gamma(d)}\int\limits_0^\infty dx\:\frac{x^{d-1}}{e^{x+z}-1}=\sum_{s=1}^\infty\frac{e^{-sx}}{s^d}\label{BE_int}
\end{equation}
were introduced, which have the properties
\begin{align}
g_s'(x)&=-g_{s-1}(x),\\
g_s(0)&=\zeta(s)\mbox{ when }s>0,
\end{align}
and $\zeta(x)$ is the Riemann zeta function. Analyzing separately small-$s$ ($s\lesssim R$) and large-$s$ ($s\gtrsim R$) parts of the remaining sum in the right hand side of Eq.~(\ref{log_sum2}), we can estimate the whole sum as $\mathcal{O}(\log N)$, thus
\begin{equation}
-\sum_\mathbf{k}\log\left(1-e^{\frac{-\mathbf{k}^2+2\mathbf{v}\cdot\mathbf{k}}R+z}\right)=\frac{N}{\zeta\left(\frac32\right)}\left(\frac{T}{T_\mathrm{c}}\right)^{3/2}g_{5/2}\left(-z-\frac{\mathbf{v}^2}R\right)+\mathcal{O}(\log N).\label{f2}
\end{equation}

The saddle-point conditions involving $z$- and $v_x$-derivatives of the logarithmic sum (\ref{f2}) need more accurate treatment because some levels $\mathbf{k}$ which host the condensate can provide large, linear in $N$, contributions to these derivatives. These levels are defined by the condition
\begin{equation}
\frac{-\mathbf{k}^2+2\mathbf{v}\cdot\mathbf{k}}R+z=0\quad\mbox{or}\quad(\mathbf{k}-\mathbf{v})^2=\mathbf{v}^2+zR
\end{equation}
of divergence of their mean occupation numbers. By the virtue of Eq.~(\ref{Jacobi_asympt}), such states provide long tails of the sum at large $t=s/R$. Denoting the set of such states as $C$, we can exclude them from the momentum summation, which results in subtracting $\sum_{\mathbf{q}\in C}e^{\frac{s(-\mathbf{q}^2+2\mathbf{v}\cdot\mathbf{q})}R}$ from the product of Jacobi functions. Subtracting also, as previously, the small-$s$ asymptotic $(\pi R/s)^{3/2}e^{\frac{s\mathbf{v}^2}R}$, we obtain the $z$-derivative of the noncondensate part of Eq.~(\ref{log_sum1}) in the form
\begin{align}
-\frac\partial{\partial z}\sum_{\mathbf{k}\notin C}\log\left(1-e^{\frac{-\mathbf{k}^2+2\mathbf{v}\cdot\mathbf{k}}R+z}\right)&=\frac{N}{\zeta(\frac32)}\left(\frac{T}{T_\mathrm{c}}\right)^{3/2}g_{3/2}\left(-z-\frac{\mathbf{v}^2}R\right)\nonumber\\
&+\sum_{s=1}^\infty e^{sz}\left[\prod_{i=x,y,z}G\left(v_i,\frac{s}R\right)-\sum_{\mathbf{q}\in C}e^{\frac{s(-\mathbf{q}^2+2\mathbf{v}\cdot\mathbf{q})}R}-\left(\frac{\pi R}s\right)^{3/2}e^{\frac{s\mathbf{v}^2}R}\right].\label{log_sum_der1}
\end{align}
The remaining sum is quickly convergent because we have subtracted both small- and large-$s$ asymptotics of the product of Jacobi functions. This sum can be estimated as $\mathcal{O}(N^{2/3})$ if we turn it to an integral.

Similarly, after differentiating the noncondensate part of Eq.~(\ref{log_sum1}) with respect to $v_x$, and subtracting the asymptotics $\sum_{\mathbf{q}\in C}(2q_x/R)e^{\frac{s(-\mathbf{q}^2+2\mathbf{v}\cdot\mathbf{q})}R}+(2v_x/R)(\pi R/s)^{3/2}e^{\frac{s\mathbf{v}^2}R}$, we obtain
\begin{align}
-\frac\partial{\partial v_x}\sum_{\mathbf{k}\notin C}\log&\left(1-e^{\frac{-\mathbf{k}^2+2\mathbf{v}\cdot\mathbf{k}}R+z}\right)=\frac2R\left\{\vphantom{\left[\prod_{i=y,z}\right]}\frac{v_xN}{\zeta(\frac32)}\left(\frac{T}{T_\mathrm{c}}\right)^{3/2}g_{3/2}\left(-z-\frac{\mathbf{v}^2}R\right)\right.\nonumber\\
&\left.+\sum_{s=1}^\infty e^{sz}\left[\frac{R}{2s}\frac{\partial G(v_x,s/R)}{\partial v_x}\prod_{i=y,z}\!\!G\left(v_i,\frac{s}R\right)-\sum_{\mathbf{q}\in C}q_xe^{\frac{s(-\mathbf{q}^2+2\mathbf{v}\cdot\mathbf{q})}R}-v_x\left(\frac{\pi R}s\right)^{3/2}e^{\frac{s\mathbf{v}^2}R}\right]\right\}.\label{log_sum_der2}
\end{align}
Here we separated analytically the $\mathcal{O}(N)$ total momentum of the thermal cloud, so the remaining quickly convergent sum in the right hand side of Eq.~(\ref{log_sum_der2}) is estimated as $\mathcal{O}(N^{2/3})$.

To calculate distribution functions, we will also need similar calculations for second derivatives of the noncondensate part of the sum (\ref{log_sum1}) with respect to $z$ and $v_x$:
\begin{align}
-\frac{\partial^2}{\partial z^2}\sum_{\mathbf{k}\notin C}\log\left(1-e^{\frac{-\mathbf{k}^2+2\mathbf{v}\cdot\mathbf{k}}R+z}\right)&=\sum_{s=1}^\infty se^{sz}\left[\prod_{i=x,y,z}G\left(v_i,\frac{s}R\right)-\sum_{\mathbf{q}\in C}e^{\frac{s(-\mathbf{q}^2+2\mathbf{v}\cdot\mathbf{q})}R}\right],\label{log_sum_der3}\\
-\frac{\partial^2}{\partial z\partial v_x}\sum_{\mathbf{k}\notin C}\log\left(1-e^{\frac{-\mathbf{k}^2+2\mathbf{v}\cdot\mathbf{k}}R+z}\right)&=\sum_{s=1}^\infty e^{sz}\left[\frac{\partial G(v_x,s/R)}{\partial v_x}\prod_{i=y,z}G\left(v_i,\frac{s}R\right)-\frac{2s}R\sum_{\mathbf{q}\in C}q_xe^{\frac{s(-\mathbf{q}^2+2\mathbf{v}\cdot\mathbf{q})}R}\right],\label{log_sum_der4}\\
-\frac{\partial^2}{\partial v_x^2}\sum_{\mathbf{k}\notin C}\log\left(1-e^{\frac{-\mathbf{k}^2+2\mathbf{v}\cdot\mathbf{k}}R+z}\right)&=\frac{2N}{R\zeta(\frac32)}\left(\frac{T}{T_\mathrm{c}}\right)^{3/2}g_{3/2}(-z)\nonumber\\
&\!\!\!\!\!\!\!\!\!\!\!\!\!\!\!\!\!\!\!\!\!\!\!\!\!\!\!\!\!\!\!\!\!\!\!\!+\sum_{s=1}^\infty \frac{e^{sz}}s\left[\frac{\partial^2G(v_x,s/R)}{\partial v_x^2}\prod_{i=y,z}G\left(v_i,\frac{s}R\right)-\frac{4s^2}{R^2}\sum_{\mathbf{q}\in C}q_x^2e^{\frac{s(-\mathbf{q}^2+2\mathbf{v}\cdot\mathbf{q})}R}-2\pi\left(\frac{\pi R}s\right)^{1/2}\right].\label{log_sum_der5}
\end{align}
If all condensate levels $\mathbf{k}\in C$ are excluded from the sums in the left hand sides, then the remaining quickly converging sums over $s$ in the right hand sides of Eqs.~(\ref{log_sum_der3}), (\ref{log_sum_der4}), and (\ref{log_sum_der5}) are estimated as $\mathcal{O}(N^{4/3})$,  $\mathcal{O}(N^{2/3})$, and  $\mathcal{O}(1)$, respectively. It is sufficient for us to neglect $-\mathbf{v}^2/R$ in the argument of $g_{3/2}$ in Eq.~(\ref{log_sum_der5}) in the range $0\leqslant v_x\leqslant1/2$ considered when we calculate the distribution functions. All higher (than the second) derivatives have simpler form because there is no need to separate the small-$s$ parts of the sums:
\begin{align}
&-\frac{\partial^{n_0+n_x+n_y+n_x}}{\partial z^{n_0}\partial v_x^{n_x}\partial v_y^{n_y}\partial v_z^{n_z}}\sum_{\mathbf{k}\notin C}\log\left(1-e^{\frac{-\mathbf{k}^2+2\mathbf{v}\cdot\mathbf{k}}R+z}\right)\nonumber\\
&=\sum_{s=1}^\infty s^{n_0-1}e^{sz}\left[\prod_{i=x,y,z}\frac{\partial^{n_i}G(v_i,s/R)}{\partial v_i^{n_i}}-\left(\frac{2s}R\right)^{n_x+n_y+n_z}\sum_{\mathbf{q}\in C}q_x^{n_x}q_y^{n_y}q_z^{n_z}e^{\frac{s(-\mathbf{q}^2+2\mathbf{v}\cdot\mathbf{q})}R}\right].\label{log_sum_der6}
\end{align}
If all condensate levels $C$ are excluded here, then the sum over $s$ quickly converges at both small and large $s$, so it can be turned to the integral and easily estimated in the large-$N$ limit.

\section{Calculation of distribution functions}

\subsection{General formulas}

Here we reproduce the general formulas for distribution functions from the main text of the paper in order to make references to them during more detailed calculations. Distribution functions $p(N_\mathbf{q})$ for occupation numbers at $\mathbf{q}=0$ or $\mathbf{q}=\mathbf{e}_x$ are given by the four-dimensional integrals [see Eq.~(13) in the main text], which is dominated in the large-$N$ limit by the Gaussian integral in the vicinity of the saddle point $(\tilde{z}(N_\mathbf{q}),\tilde{\mathbf{v}}(N_\mathbf{q}))$, whose location \emph{depends on} $N_\mathbf{q}$:
\begin{equation}
p(N_\mathbf{q})\sim \frac1{\sqrt{H}}e^{\tilde{f}(N_\mathbf{q},\tilde{z}(N_\mathbf{q}),\tilde{\mathbf{v}}(N_\mathbf{q}))}.\label{p4}
\end{equation}
Here we can write the function $\tilde{f}$ [Eq.~(15) from the main text]
\begin{equation}
\tilde{f}(N_\mathbf{q},z,\mathbf{v})=h(z,\mathbf{v})+zN_\mathbf{q}+\frac{(-\mathbf{q}^2+2\mathbf{v}\cdot\mathbf{q})N_\mathbf{q}}R\label{g2}
\end{equation}
in terms of
\begin{equation}
h(z,\mathbf{v})=-zN-\frac{2\mathbf{v}\cdot\mathbf{Q}}R-\sum_{\mathbf{k}\neq\mathbf{q}}\log\left(1-e^{\frac{-\mathbf{k}^2+2\mathbf{v}\cdot\mathbf{k}}R+z}\right)\label{h}.
\end{equation}
The Hessian determinant
\begin{equation}
H=\left.\left\{\frac{\partial^2h}{\partial z^2}\frac{\partial^2h}{\partial v_x^2}-\left(\frac{\partial^2h}{\partial z\partial v_x}\right)^2\right\}\right|\strut_{z=\tilde{z}(N_\mathbf{q})\atop\mathbf{v}=\tilde{\mathbf{v}}(N_\mathbf{q})}\label{Hessian}
\end{equation}
of both $\tilde{f}$ and $h$, calculated at the saddle point, provides to Eq.~(\ref{p4}) subleading contribution with respect to the leading-order exponential term. It will be taken into account in the following to obtain refined expressions for location of the distribution maxima.

The saddle point location $(\tilde{z}(N_\mathbf{q}),\tilde{\mathbf{v}}(N_\mathbf{q}))$ is defined by the stationarity conditions
\begin{equation}
\frac{\partial\tilde{f}}{\partial z}=\frac{\partial h}{\partial z}+N_\mathbf{q}=0,\qquad
\frac{\partial\tilde{f}}{\partial\mathbf{v}}=\frac{\partial h}{\partial\mathbf{v}}+\frac{2\mathbf{q}N_\mathbf{q}}R=0.\label{gder4}
\end{equation}
The maximum of distribution function (\ref{p4}) is \emph{formally} achieved at some particle number $N_\mathbf{q}=\bar{N}_\mathbf{q}$, where
\begin{align}
\left.\frac{d}{dN_\mathbf{q}}\left(\tilde{f}-\frac12\log H\right)\right|\strut_{z=\tilde{z}(N_\mathbf{q})\atop\mathbf{v}=\tilde{\mathbf{v}}(N_\mathbf{q})}=\frac{-\mathbf{q}^2+2\tilde{\mathbf{v}}(N_\mathbf{q})\cdot\mathbf{q}}R+\tilde{z}(N_\mathbf{q})-\frac12\frac{d\log H}{dN_\mathbf{q}}=0.\label{max_cond}
\end{align}
We have used the stationarity conditions (\ref{gder4}) in derivation of this formula. The \emph{actual} location of the maximum coincides with $\bar{N}_\mathbf{q}$ if $\bar{N}_\mathbf{q}>0$, so we can perform Gaussian expansion of $p(N_\mathbf{q})$ around $\bar{N}_\mathbf{q}$. Otherwise, if $\bar{N}_\mathbf{q}<0$, only a right tail of this Gaussian distribution will be present in the physical region $N_\mathbf{q}\geqslant0$, so in the large-$N$ limit we can leave only its exponential asymptotic.

Thus, in the case $\bar{N}_\mathbf{q}>0$, we expand the distribution (\ref{p4}) around the maximum $N_\mathbf{q}=\bar{N}_\mathbf{q}$ in the Gaussian form
\begin{equation}
p(N_\mathbf{q})\sim \exp\left\{\frac12\left.\frac{d^2\tilde{f}}{dN_\mathbf{q}^2}\right|\strut_{z=\tilde{z}(\bar{N}_\mathbf{q})\atop\mathbf{v}=\tilde{\mathbf{v}}(\bar{N}_\mathbf{q})}(N_\mathbf{q}-\bar{N}_\mathbf{q})^2\right\}.\label{p5}
\end{equation}
To calculate the second derivative $d^2\tilde{f}/dN_\mathbf{q}^2$ at the saddle point we use Eqs.~(\ref{g2})--(\ref{h}) together with the stationarity conditions (\ref{gder4}) and their derivatives with respect to $N_\mathbf{q}$. The result is
\begin{equation}
\left.\frac{d^2\tilde{f}}{dN_\mathbf{q}^2}\right|\strut_{z=\tilde{z}(\bar{N}_\mathbf{q})\atop\mathbf{v}=\tilde{\mathbf{v}}(\bar{N}_\mathbf{q})}=\frac{d\tilde{z}}{dN_\mathbf{q}}+\frac{2\mathbf{q}}R\cdot\frac{d\tilde{\mathbf{v}}}{dN_\mathbf{q}}.\label{gder6}
\end{equation}
On the other hand, differentiating Eq.~(\ref{gder4}) by $N_\mathbf{q}$ and taking into account that the mixed second derivatives $\partial^2h/\partial z\partial v_{y,z}$ and $\partial^2h/\partial v_i\partial v_j$ (at $i\neq j$) vanish at the saddle point since $\tilde{v}_y=\tilde{v}_z=0$, we obtain
\begin{align}
\frac{d\tilde{z}}{dN_\mathbf{q}}&=\frac1H\left(\frac{2q_x}R\frac{\partial^2h}{\partial z\partial v_x}-\frac{\partial^2h}{\partial v_x^2}\right),\label{spder1}\\
\frac{d\tilde{\mathbf{v}}}{dN_\mathbf{q}}&=\frac{\mathbf{e}_x}{H}\left(-\frac{2q_x}R\frac{\partial^2h}{\partial z^2}+\frac{\partial^2h}{\partial z\partial v_x}\right).\label{spder2}
\end{align}
Substituting these expressions to Eq.~(\ref{gder6}) and then to Eq.~(\ref{p5}), we get
\begin{align}
p(N_\mathbf{q})
\sim\exp&\left\{-\left(\frac{2q_x^2}{R^2}\frac{\partial^2h}{\partial z^2}-\frac{2q_x}R\frac{\partial^2h}{\partial z\partial v_x}+\frac12\frac{\partial^2h}{\partial v_x^2}\right)\frac{(N_\mathbf{q}-\bar{N}_\mathbf{q})^2}H\right\}.\label{p6}
\end{align}
In the formulas (\ref{spder1})--(\ref{p6}), all derivatives of $h$ should be evaluated at the saddle point $(\tilde{z}(\bar{N}_\mathbf{q}),\tilde{\mathbf{v}}(\bar{N}_\mathbf{q}))$.

In the case $\bar{N}_\mathbf{q}<0$, we expand $\tilde{f}$ in Eq.~(\ref{p4}) up to the linear term around $N_\mathbf{q}=0$ using the maximum condition (\ref{max_cond}) with the subleading terms neglected:
\begin{equation}
p(N_\mathbf{q})
\sim\exp\left\{\left[\tilde{z}(0)+\frac{-\mathbf{q}^2+2\tilde{\mathbf{v}}(0)\cdot\mathbf{q}}R\right]N_\mathbf{q}\right\}.\label{p7}
\end{equation}
Thus we have obtained explicit formulas for distribution functions both in Gaussian (\ref{p6}) and exponential (\ref{p7}) regimes.

\subsection{Distribution functions in the normal state, $T>T_\mathrm{c}$}

To find the distribution functions $p(N_\mathbf{q})$ for occupation numbers $N_\mathbf{q}$ of the 0th ($\mathbf{q}=0$) and 1st ($\mathbf{q}=\mathbf{e}_x$) levels, we use Eqs.~(\ref{g2})--(\ref{h}), and (\ref{log_sum_der1})--(\ref{log_sum_der2}) with $C=\emptyset$ to write the stationarity conditions (\ref{gder4}) in the leading ($\sim N$) order as
\begin{align}
\frac{\partial\tilde{f}}{\partial z}&=-N+N_\mathbf{q}+\frac{N}{\zeta(\frac32)}\left(\frac{T}{T_\mathrm{c}}\right)^{3/2}g_{3/2}(-z)=0,\label{p_norm_st1}\\
\frac{\partial\tilde{f}}{\partial v_x}&=\frac2R\left\{-Q+q_xN_\mathbf{q}+\frac{v_xN}{\zeta(\frac32)}\left(\frac{T}{T_\mathrm{c}}\right)^{3/2}g_{3/2}(-z)\right\}=0.\label{p_norm_st2}
\end{align}
For both levels, the last term $d\log H/dN_\mathbf{q}=\mathcal{O}(N^{-1})$ in the maximum condition (\ref{max_cond}) can be disregarded, so we obtain $\tilde{z}(N_\mathbf{q})=\mathcal{O}(N^{-2/3})$. Given such small $\tilde{z}$ which result in $g_{3/2}(-\tilde{z})\approx\zeta(\frac32)$, Eq.~(\ref{p_norm_st1}) yields the negative formal locations of the distribution maxima $\bar{N}_\mathbf{q}\approx N[1-(T/T_\mathrm{c})^{3/2}]<0$, since $T>T_\mathrm{c}$. This indicates the absence of condensate on both levels, thus we apply the exponential form (\ref{p7}) of the distribution function near the actual maximum point $N_\mathbf{q}=0$ to get:
\begin{align}
p(N_0)&\sim\exp\{\tilde{z}(0)N_0\},\\
p(N_1)&\sim\exp\left\{\left[\tilde{z}(0)+\frac{2\tilde{v}_x(0)-1}R\right]N_1\right\}.
\end{align}
Here the dimensionless chemical potential $\tilde{z}(0)$ is found from Eq.~(\ref{p_norm_st1}) taken at $N_\mathbf{q}=0$,
\begin{equation}
\frac1{\zeta(\frac32)}\left(\frac{T}{T_\mathrm{c}}\right)^{3/2}g_{3/2}(-\tilde{z}(0))=1,\label{zt_norm}
\end{equation}
and $\tilde{v}_x(0)=Q/N$ from Eq.~(\ref{p_norm_st2}).

\subsection{Distribution functions for 0th level at $T<T_\mathrm{c}$}

For the Bose-condensed phase, we need more elaborate treatment to obtain subleading contributions to locations of distribution maxima. In the unfragmented BEC phase $T<T_\mathrm{c}$, $Q<Q_0$, we use Eqs.~(\ref{g2})--(\ref{h}), and (\ref{log_sum_der1})--(\ref{log_sum_der2}) with the condensate level $C=\{0\}$ to write the stationarity conditions (\ref{gder4}) as
\begin{align}
\frac{\partial\tilde{f}}{\partial z}&=-N+N_0+\frac{N}{\zeta(\frac32)}\left(\frac{T}{T_\mathrm{c}}\right)^{3/2}g_{3/2}(-z)+RI_1(v_x,Rz)+o(N^{2/3})=0,\label{p0_BEC1_st1}\\
\frac{\partial\tilde{f}}{\partial v_x}&=\frac2R\left\{-Q+\frac{v_xN}{\zeta(\frac32)}\left(\frac{T}{T_\mathrm{c}}\right)^{3/2}g_{3/2}(-z)+\mathcal{O}(N^{2/3})\right\}=0.\label{p0_BEC1_st2}
\end{align}
Here the sum over $s$ in Eq.~(\ref{log_sum_der1}) was transformed to the integral
\begin{equation}
I_1(v_x,\alpha)=\int\limits_0^\infty dt\:e^{\alpha t}[G(v_x,t)G^2(0,t)-1-(\pi/t)^{3/2}]
\end{equation}
over $t=s/R$ with the subleading error $o(R)=o(N^{2/3})$ to retain the $N^{2/3}$ term $RI_1(v_x,Rz)$ in Eq.~(\ref{p0_BEC1_st1}). In order to employ the condition (\ref{max_cond}) for the distribution maximum $N_0=\bar{N}_0$, we find the second derivatives of the function (\ref{h}) using Eqs.~(\ref{log_sum_der3})--(\ref{log_sum_der5}) with $C=\{0\}$: 
\begin{align}
&\frac{\partial^2h}{\partial z^2}=R^2J_1(\tilde{v}_x(N_0),R\tilde{z}(N_0))+o(N^{4/3}),\label{p0_BEC1_sder1}\\
&\frac{\partial^2h}{\partial z\partial v_x}=\mathcal{O}(N^{2/3}),\label{p0_BEC1_sder2}\\
&\frac{\partial^2h}{\partial v_x^2}=\frac{2N}R\left(\frac{T}{T_\mathrm{c}}\right)^{3/2}+\mathcal{O}(1).\label{p0_BEC1_sder3}
\end{align}
In the last formula, we have taken into account that $\tilde{z}$ is as small as $N^{-2/3}$ or even smaller, as will be confirmed below. Also the sum over $s$ in Eq.~(\ref{log_sum_der3}) was transformed into the integral
\begin{equation}
J_n(v_x,\alpha)=\int\limits_0^\infty dt\:t^ne^{\alpha t}[G(v_x,t)G^2(0,t)-1]\label{J_n}
\end{equation}
over $t=s/R$ with the subleading error $o(N^{4/3})$ appearing in the right-hand side of Eq.~(\ref{p0_BEC1_sder1}). The Hessian determinant
\begin{equation}
H=2NR\left(\frac{T}{T_\mathrm{c}}\right)^{3/2}J_1(\tilde{v}_x(N_0),R\tilde{z}(N_0))+\mathcal{O}(N^{4/3}),
\end{equation}
calculated from the second derivatives (\ref{p0_BEC1_sder1})--(\ref{p0_BEC1_sder3}), provides the correction term
\begin{equation}
\frac{d\log H}{dN_0}=-\frac{J_2(\tilde{v}_x(N_0),R\tilde{z}(N_0))}{RJ_1^2(\tilde{v}_x(N_0),R\tilde{z}(N_0))}\label{p0_H_der}
\end{equation}
to the distribution maximum condition (\ref{max_cond}). Here we have taken into account the weak [$d\tilde{v}_x/dN_0\sim N^{-1}$, see Eq.~(\ref{spder2})] dependence of $\tilde{v}_x$ on $N_0$. Defining the variable $\alpha_0=R\tilde{z}(\bar{N}_0)$, we obtain the maximum condition in the form of equation 
\begin{equation}
\alpha_0+\frac{J_2(\tilde{v}_x(\bar{N}_0),\alpha_0)}{2J_1^2(\tilde{v}_x(\bar{N}_0),\alpha_0)}=0\label{p0_BEC1_max_cond}
\end{equation}
for $\alpha_0$ which can be solved numerically at given $\tilde{v}_x(\bar{N}_0)$. Relating $\tilde{z}(\bar{N}_0)=\alpha_0/R$ to the numerically found $\alpha_0$,  we obtain from the stationarity conditions (\ref{p0_BEC1_st1})-- (\ref{p0_BEC1_st2}):
\begin{align}
&\bar{N}_0=N\left[1-\left(\frac{T}{T_\mathrm{c}}\right)^{3/2}\right]-RI_1(v_x,\alpha_0)+o(N^{2/3}),\label{p0_BEC1_n0}\\
&\tilde{v}_x(\bar{N}_0)=\frac{Q}{2Q_0}+\mathcal{O}(N^{-1/3}).\label{p0_BEC1_vx}
\end{align}
The formula (\ref{p0_BEC1_vx}) for rapidity can be used in the equation (\ref{p0_BEC1_max_cond}) for $\alpha_0$. Calculating the width of the distribution (\ref{p6}) using Eqs.~(\ref{p0_BEC1_sder1})--(\ref{p0_BEC1_sder3}), we obtain the final formula
\begin{equation}
p(N_0)\sim\exp\left\{-\frac{(N_0-\bar{N}_0)^2}{2R^2J_1(Q/2Q_0,\alpha_0)}\right\}.\label{p0_BEC1_distr}
\end{equation}

In the fragmented BEC state $T<T_\mathrm{c}$, $Q_0<Q\leqslant N/2$, we need to single out the occupation
\begin{equation}
\tilde\nu_1=\frac1{e^{\frac{1-2\tilde{v}_x(N_0)}R-\tilde{z}(N_0)}-1}
\end{equation}
of the 1st level which is of the order of $N$. The second derivatives of the function (\ref{h}) calculated using Eqs.~(\ref{log_sum_der3})--(\ref{log_sum_der5}) with exclusion of both condensate levels $C=\{0,\mathbf{e}_x\}$, are:
\begin{align}
&\frac{\partial^2h}{\partial z^2}=\tilde\nu_1^2+\mathcal{O}(N^{4/3}),\label{hder1_BEC2}\\
&\frac{\partial^2h}{\partial z\partial v_x}=\frac{2\tilde\nu_1^2}R+\mathcal{O}(N^{2/3}),\\
&\frac{\partial^2h}{\partial v_x^2}=\frac{4\tilde\nu_1^2}{R^2}+\frac{2N}R\left(\frac{T}{T_\mathrm{c}}\right)^{3/2}+\mathcal{O}(1).\label{hder2_BEC2}
\end{align}
Here we have taken into account that $\tilde{z}=\mathcal{O}(N^{-1})$, as will be confirmed below. The subtle point here is that the leading-order $N^{8/3}$ terms in the Hessian determinant (\ref{Hessian}) cancel each other, so that only the subleading $N^{7/3}$ result remains:
\begin{equation}
H=\frac{2N\tilde\nu_1^2}R\left(\frac{T}{T_\mathrm{c}}\right)^{3/2}+\mathcal{O}(N^2).\label{Hessian_BEC1}
\end{equation}
In other words, the matrix of second derivatives is almost degenerate.

The correction term $d\log H/dN_0\approx-2/\tilde\nu_1$ in the condition for distribution maximum (\ref{max_cond}) can be calculated from  Eqs.~(\ref{spder1})--(\ref{spder2}) and (\ref{hder1_BEC2})--(\ref{Hessian_BEC1}) with taking into account that $\tilde\nu_1\approx[-\tilde{z}(N_0)+(1-2\tilde{v}_x(N_0))/R]^{-1}$. Thus we obtain the saddle point location at the distribution maximum $N_0=\bar{N}_0$:
\begin{align}
\tilde{z}(\bar{N}_0)&=-\frac1{Q-Q_0}+\mathcal{O}(N^{-4/3}),\label{p0_BEC2_zt}\\
\tilde{v}_x(\bar{N}_0)&=\frac12+\mathcal{O}(N^{-2/3}),\label{p0_BEC2_vt}
\end{align}
and mean occupation of the 1st level is
\begin{equation}
\tilde\nu_1=Q-Q_0+\mathcal{O}(N^{2/3}).\label{p0_BEC2_nu1}
\end{equation}
Now we can derive the accurate form of stationarity conditions (\ref{gder4}) at these parameters using Eqs.~(\ref{g2})--(\ref{h}), and (\ref{log_sum_der1})--(\ref{log_sum_der2}) with the condensate levels $C=\{0,\mathbf{e}_x\}$:
\begin{align}
\frac{\partial\tilde{f}}{\partial z}&=-N+N_0+\tilde\nu_1+N\left(\frac{T}{T_\mathrm{c}}\right)^{3/2}-RI_2+o(N^{2/3})=0,\label{p0_BEC2_st1}\\
\frac{\partial\tilde{f}}{\partial v_x}&=\frac2R\left\{-Q+\tilde\nu_1+\frac{N}2\left(\frac{T}{T_\mathrm{c}}\right)^{3/2}-\frac{RI_2}2+o(N^{2/3})\right\}=0.\label{p0_BEC2_st2}
\end{align}
Here we transformed the sums over $s$ in Eqs.~(\ref{log_sum_der1})--(\ref{log_sum_der2}) to the integral
\begin{equation}
I_2=-\int\limits_0^\infty dt\left[G\left(\frac12,t\right)G^2(0,t)-2-\left(\frac\pi{t}\right)^{3/2}\right]\approx6.375\label{I3}
\end{equation}
over $t=s/R$ with the subleading error $o(N^{2/3})$ and used the property
\begin{equation}
\frac1{2t}\left.\frac{\partial G(v_x,t)}{\partial v_x}\right|_{v_x=1/2}=\frac12G\left(\frac12,t\right)\label{Jacobi_prop}
\end{equation}
of the Jacobi functions (\ref{Jacobi}). Solving equations (\ref{p0_BEC2_st1})--(\ref{p0_BEC2_st2}), we obtain the location of distribution maximum
\begin{equation}
\bar{N}_0=N\left[1-\left(\frac{T}{T_\mathrm{c}}\right)^{3/2}\right]-(Q-Q_0)+\frac{RI_2}2+o(N^{2/3})
\end{equation}
calculated with taking into account the subleading $R\sim N^{2/3}$ term. The final form of distribution function (\ref{p6}), obtained using Eqs.~(\ref{hder1_BEC2})--(\ref{Hessian_BEC1}) and (\ref{p0_BEC2_zt})--(\ref{p0_BEC2_nu1}), reads
\begin{equation}
p(N_0)\sim\exp\left\{-\frac{(N_0-\bar{N}_0)^2}{NR(T/T_\mathrm{c})^{3/2}}\right\}.\label{p0_BEC2_distr}
\end{equation}

\subsection{Distribution functions for 1st level at $T<T_\mathrm{c}$}

For both fragmented and unfragmented BEC states at $T<T_\mathrm{c}$, we expect condensation at the 0th level, so we need to single out the occupation
\begin{equation}
\tilde\nu_0=\frac1{e^{-\tilde{z}(N_1)}-1}
\end{equation}
of this level, which is of the order of $N$. The stationarity conditions (\ref{gder4}), obtained using Eqs.~(\ref{g2})--(\ref{h}) and (\ref{log_sum_der1})--(\ref{log_sum_der2}) with exclusion of both condensate levels $C=\{0,\mathbf{e}_x\}$, are:
\begin{align}
\frac{\partial\tilde{f}}{\partial z}&=-N+\tilde\nu_0+N_1+\frac{N}{\zeta(\frac32)}\left(\frac{T}{T_\mathrm{c}}\right)^{3/2}g_{3/2}(-z)\nonumber\\
&+R\int\limits_0^\infty dt\:\left[G(v_x,t)G^2(0,t)-1-e^{\frac{s(2v_x-1)}R}-(\pi/t)^{3/2}\right]+o(N^{2/3})=0,\label{p1_BEC_st1}\\
\frac{\partial\tilde{f}}{\partial v_x}&=\frac2R\left\{-Q+N_1+\frac{v_xN}{\zeta(\frac32)}\left(\frac{T}{T_\mathrm{c}}\right)^{3/2}g_{3/2}(-z)\right.\nonumber\\
&\qquad\quad\left.+R\int\limits_0^\infty dt\:\left[\frac1{2t}\frac{\partial G(v_x,t)}{\partial v_x}G^2(0,t)-e^{\frac{s(2v_x-1)}R}-v_x(\pi/t)^{3/2}\right]+o(N^{2/3})\vphantom{\left(\frac{T}{T_\mathrm{c}}\right)^{3/2}}\right\}=0.\label{p1_BEC_st2}
\end{align}
Here the sums over $s$ were transformed to the integrals over $t=s/R$ with subleading errors $o(N^{2/3})$.

Let us take into account that $\tilde{z}$ at the saddle point is as small as $N^{-2/3}$ or smaller, as will be confirmed below by the distribution maximum condition. It allows us to estimate the second derivatives of the function (\ref{h}) calculated from Eqs.~(\ref{log_sum_der3})--(\ref{log_sum_der5}) with $C=\{0,\mathbf{e}_x\}$:
\begin{align}
&\frac{\partial^2h}{\partial z^2}=\tilde\nu_0^2+\mathcal{O}(N^{4/3}),\label{hder1_BEC3}\\
&\frac{\partial^2h}{\partial z\partial v_x}=\mathcal{O}(N^{2/3}),\\
&\frac{\partial^2h}{\partial v_x^2}=\frac{2N}R\left(\frac{T}{T_\mathrm{c}}\right)^{3/2}+\mathcal{O}(1).\label{hder2_BEC3}
\end{align}
The Hessian determinant (\ref{Hessian}) is
\begin{equation}
H=\frac{2N\tilde\nu_0^2}R\left(\frac{T}{T_\mathrm{c}}\right)^{3/2}+\mathcal{O}(N^2),\label{Hessian_BEC3}
\end{equation}
and $d\log H/dN_1\approx(2/\tilde\nu_0)(d\tilde\nu_0/dN_1)\approx-(2/\tilde{z})(d\tilde{z}/dN_1)$, since $\tilde\nu_0\approx-1/\tilde{z}$. Using Eqs.~(\ref{spder1}) and (\ref{hder1_BEC3})--(\ref{hder2_BEC3}), we obtain $d\log H/dN_1=2\tilde{z}+\mathcal{O}(N^{-4/3})$. Substituting it to the maximum condition (\ref{max_cond}), we obtain the rapidity
\begin{equation}
\tilde{v}_x(\bar{N}_1)=\frac12+\mathcal{O}(N^{-2/3}).\label{p1_BEC2_vt}
\end{equation}
This equation, in combination with Eq.~(\ref{p1_BEC_st2}), yields the formal maximum location of the distribution $\bar{N}_1=Q-Q_0+\mathcal{O}(N^{2/3})$. We conclude that, in the unfragmented BEC phase $T<T_\mathrm{c}$, $Q<Q_0$, the maximum point is negative, so we assume the actual location of the maximum at $N_1=0$. Substituting it to Eqs.~(\ref{p1_BEC_st1})--(\ref{p1_BEC_st2}), we obtain
\begin{align}
\tilde{z}(0)&=-\frac1{N[1-(T/T_\mathrm{c})^{3/2}]}+\mathcal{O}(N^{-4/3}),\label{p1_BEC1_zt}\\
\tilde{v}_x(0)&=Q/2Q_0+\mathcal{O}(N^{-1/3}).\label{p1_BEC1_vt}
\end{align}
For the distribution function $p(N_1)$, we would like to derive more accurate approximation than the exponential one (\ref{p7}), because the latter produces large numerical errors. To do it, we retain, besides the linear-$N_1$ term in the exponent of Eq.~(\ref{p4}), the quadratic term too, which has the same form as in Eq.~(\ref{p6}), but should be calculated at the actual maximum point location $N_1=0$. Using Eqs.~(\ref{p1_BEC1_zt})--(\ref{p1_BEC1_vt}) and (\ref{hder1_BEC3})--(\ref{Hessian_BEC3}), we obtain the coefficient at this quadratic term:
\begin{equation}
\frac1H\left(\frac{2q_x^2}{R^2}\frac{\partial^2h}{\partial z^2}-\frac{2q_x}R\frac{\partial^2h}{\partial z\partial v_x}+\frac12\frac{\partial^2h}{\partial v_x^2}\right)\approx\frac1{NR(T/T_c)^{3/2}},
\end{equation}
and the final distribution function, with both linear and quadratic terms, reads:
\begin{equation}
p(N_1)\sim\exp\left\{-\frac{(Q_0-Q)N_1}{Q_0R}-\frac{N_1^2}{NR(T/T_c)^{3/2}}\right\}.
\end{equation}

In the fragmented BEC phase $T<T_\mathrm{c}$, $Q_0<Q\leqslant N/2$, we can use Eqs.~(\ref{p1_BEC2_vt}) and (\ref{Jacobi_prop}) to recast the stationarity conditions (\ref{p1_BEC_st1})--(\ref{p1_BEC_st2}) in the form
\begin{align}
-N+\tilde\nu_0+N_1+N\left(\frac{T}{T_\mathrm{c}}\right)^{3/2}-RI_2+o(N^{2/3})=0,\label{p1_BEC2_st1}\\
-Q+N_1+Q_0-\frac{RI_2}2+o(N^{2/3})=0,\label{p1_BEC2_st2}
\end{align}
where the integral $I_2$ is defined in Eq.~(\ref{I3}). Since these conditions are valid at the maximum point $N_1=\bar{N}_1$ of the distribution function, we immediately obtain its location from Eq.~(\ref{p1_BEC2_st2}):
\begin{equation}
\bar{N}_1=Q-Q_0+\frac{RI_2}2+o(N^{2/3}).
\end{equation}
Using Eqs.~(\ref{p6}) and (\ref{hder1_BEC3})--(\ref{Hessian_BEC3}), we can write the final distribution function
\begin{equation}
p(N_1)\sim\exp\left\{-\frac{(N_1-\bar{N}_1)^2}{NR(T/T_\mathrm{c})^{3/2}}\right\}.
\end{equation}

\section{Effect of momentum fixation}

In canonical ensemble without fixation of the total momentum, partition and distribution functions can be found by similar saddle-point integration as in Eqs.~(12)--(15) of the main text, but without integration over $\mathbf{v}$:
\begin{align}
Z_\mathrm{c}=&\frac1{2\pi i}\int\limits_{-\pi i}^{\pi i}dz\:e^{f_\mathrm{c}(z)},\label{Z_c}\\
p_\mathrm{c}(N_\mathbf{q})=&\frac1{2\pi iZ_\mathrm{c}}\int\limits_{-\pi i}^{\pi i}dz\:e^{\tilde{f}_\mathrm{c}(N_\mathbf{q},z)},\label{p_c}
\end{align}
where
\begin{align}
f_\mathrm{c}(z)&=-zN-\sum_\mathbf{k}\log\left(1-e^{-\frac{\mathbf{k}^2}R+z}\right),\label{f_c}\\
\tilde{f}_\mathrm{c}(N_\mathbf{q},z)&=-z(N-N_\mathbf{q})-\frac{\mathbf{q}^2N_\mathbf{q}}R-\sum_{\mathbf{k}\neq\mathbf{q}}\log\left(1-e^{-\frac{\mathbf{k}^2}R+z}\right).\label{g_c}
\end{align}
Comparing these expressions with those [Eqs.~(12)--(15)], where the total momentum is fixed to be zero, $\mathbf{Q}=0$, we note that values of the exponents in the saddle-point integrals are the same: $f_\mathrm{c}(z)=f(z,\mathbf{v})|_{\mathbf{v}=0}$, $\tilde{f}_\mathrm{c}(N_\mathbf{q},\tilde{z})=\tilde{f}(N_\mathbf{q},\tilde{z},\tilde{\mathbf{v}})|_{\tilde{\mathbf{v}}=0}$. Since with $\mathbf{Q}=0$ the saddle-point conditions $\partial f/\partial\mathbf{v}=0$ and $\partial\tilde{f}/\partial\mathbf{v}=0$ are satisfied at $\mathbf{v}=0$, we conclude that the partition $Z$ and distribution  $p(N_\mathbf{q})$ functions calculated with the fixed total momentum $\mathbf{Q}=0$ differ from those calculated in canonical ensemble only by extra integration over $\mathbf{v}$ around the saddle point $(z_0,\mathbf{v}=0)$ or $(\tilde{z}(N_\mathbf{q}),\tilde{\mathbf{v}}=0)$. This three-dimensional integration gives rise to additional pre-exponential factors in $Z$ and $p(N_\mathbf{q})$.

Namely, the saddle-point expressions for partition function $Z$ at fixed $\mathbf{Q}=0$ and for its counterpart $Z_\mathrm{c}$ in canonical ensemble differ as
\begin{equation}
Z\sim\frac1{R^3\sqrt{H}}e^{f(z_0,0)},\qquad Z_\mathrm{c}\sim\frac1{\sqrt{H_\mathrm{c}}}e^{f_\mathrm{c}(z_0)},\label{Z_vs_Z_c}
\end{equation}
where the exponents $f(z_0,0)$ and $f_\mathrm{c}(z_0)$ are the same, and the Hessian determinants are
\begin{equation}
H_\mathrm{c}=-\frac{\partial^2}{\partial z^2}\sum_\mathbf{k}\log\left(1-e^{-\frac{\mathbf{k}^2}R+z}\right),\qquad
H=H_\mathrm{c}\prod_{i=x,y,z}\left\{-\frac{\partial^2}{\partial v_i^2}\left.\sum_\mathbf{k}\log\left(1-e^{\frac{-\mathbf{k}^2+2\mathbf{v}\cdot\mathbf{k}}R+z}\right)\right|_{\mathbf{v}=0}\right\}.
\end{equation}
Both at $T>T_\mathrm{c}$, when the condensate is absent, and at $T<T_\mathrm{c}$, when we need to separate the condensate level $C=\{0\}$ in the formula (\ref{log_sum_der5}) for the second derivative, we obtain the same result
\begin{equation}
-\frac{\partial^2}{\partial v_i^2}\left.\sum_\mathbf{k}\log\left(1-e^{\frac{-\mathbf{k}^2+2\mathbf{v}\cdot\mathbf{k}}R+z}\right)\right|_{\mathbf{v}=0}=\frac{2N}{R\zeta(\frac32)}\left(\frac{T}{T_\mathrm{c}}\right)^{3/2}g_{3/2}(-z)+\mathcal{O}(1)\sim N^{1/3},
\end{equation}
because the condensate contribution to this derivative vanishes at $\mathbf{v}=0$. Thus the ratio of Hessian determinants $H/H_\mathrm{c}$ is linear in $N$, and the ratio of partition functions (\ref{Z_vs_Z_c}) is $Z/Z_\mathrm{c}\sim N^{-5/2}$. In result, the difference of free energies at fixed $\mathbf{Q}=0$ and in canonical ensemble,
\begin{equation}
F|_{\mathbf{Q}=0}-F_\mathrm{c}=-T\log\frac{Z|_{\mathbf{Q}=0}}{Z_\mathrm{c}}\sim\log N,\label{F_diff}
\end{equation}
is subextensive and becomes negligible in comparison with the free energy itself. However this result is valid only at not too small temperatures, when $T/T_\mathrm{c}\gg N^{-2/3}$, otherwise the saddle-point approximation becomes poorly applicable. Specifically, in the latter case the difference (\ref{F_diff}) of free energies becomes comparable with the free energy itself. Indeed, according to Eq.~(37) from the main text, the leading-order $\mathcal{O}(N)$ part of $F|_{\mathbf{Q}=0}$ becomes of the order of 1 at $T/T_\mathrm{c}\sim N^{-2/3}$, so the remaining $\mathcal{O}(\log N)$ terms in Eq.~(37) and the difference  (\ref{F_diff}) cannot be neglected as subleading corrections.

To compare the distribution functions $p(N_0)$ and $p_\mathrm{c}(N_0)$, which are calculated, respectively, at fixed momentum $\mathbf{Q}=0$ and in canonical ensemble without momentum fixation, we heed to analyze their pre-exponential factors (inverse square roots of Hessian determinants), which affect the distribution maximum condition (\ref{max_cond}). In the case of fixed momentum $\mathbf{Q}=0$, the Hessian determinant $H=(\partial^2h/\partial z^2)(\partial^2h/\partial v_x^2)(\partial^2h/\partial v_y^2)(\partial^2h/\partial v_z^2)$ should include, by the symmetry, all three derivatives $\partial^2h/\partial v_i^2$, which are equal to each other, thus the maximum condition for occupation of the 0th level reads
\begin{equation}
\tilde{z}(N_0)-\frac12\frac{d\log H}{dN_0}=\tilde{z}(N_0)-\left\{\frac12\frac{\partial^3h/\partial z^3}{\partial^2h/\partial z^2}+\frac32\frac{\partial^3h/\partial z\partial v_x^2}{\partial^2h/\partial v_x^2}\right\}\frac{d\tilde{z}(N_0)}{dN_0}=0.\label{max_cond_fixed}
\end{equation}
Here we have calculated the derivative by $N_0$ taking into account that $d\tilde{v}_x(N_0)/dN_0=0$, see Eq.~(\ref{spder2}). Note that the $\partial^3h/\partial z\partial v_x^2$ term was neglected in the previous section when we calculated the derivative of Hessian determinant in Eq.~(\ref{p0_H_der}), because it provided the subleading contribution with respect to the $\partial^3h/\partial z^3$ term, but here we retain it. Without momentum fixation, the Hessian determinant $H_\mathrm{c}=\partial^2h/\partial z^2$ does not contain $v_i$-derivatives, so the maximum condition (\ref{max_cond}) in canonical ensemble is
\begin{equation}
\tilde{z}(N_0)-\frac12\frac{d\log H_\mathrm{c}}{dN_0}=\tilde{z}(N_0)-\frac12\frac{\partial^3h/\partial z^3}{\partial^2h/\partial z^2}\frac{d\tilde{z}(N_0)}{dN_0}=0.\label{max_cond_can}
\end{equation}
If the maximum locations $\bar{N}_0$ and $\bar{N}_{0\mathrm{c}}$ found from, respectively, Eqs.~(\ref{max_cond_fixed}) and (\ref{max_cond_can}), differ by relatively small amount, then we can expand Eq.~(\ref{max_cond_can}) around $\bar{N}_0$ and subtract from Eq.~(\ref{max_cond_fixed}) to obtain
\begin{equation}
(\bar{N}_0-\bar{N}_{0\mathrm{c}})\frac{d}{dN_0}\left\{\tilde{z}(N_0)-\frac12\frac{\partial^3h/\partial z^3}{\partial^2h/\partial z^2}\frac{d\tilde{z}}{dN_0}\right\}=\frac32\frac{\partial^3h/\partial z\partial v_x^2}{\partial^2h/\partial v_x^2}\frac{d\tilde{z}}{dN_0},\label{can_z_diff}
\end{equation}
where all functions are taken at $z=\tilde{z}(N_0)$, $N_0=\bar{N}_0$. Differentiating the first stationarity condition (\ref{gder4}) by $N_0$ once and twice, we obtain
\begin{equation}
\frac{d\tilde{z}}{dN_0}=-\frac1{\partial^2h/\partial z^2},\qquad\frac{d^2\tilde{z}}{dN_0^2}=-\frac{\partial^3h/\partial z^3}{(\partial^2h/\partial z^2)^3}\label{z_der_saddle}
\end{equation}
at the saddle point $z=\tilde{z}(N_0)$. Using Eq.~(\ref{z_der_saddle}) allows to simplify the $N_0$-derivative in Eq.~(\ref{can_z_diff}), and the result is
\begin{equation}
\bar{N}_0-\bar{N}_{0\mathrm{c}}=\frac{\frac32\frac{\partial^3h/\partial z\partial v_x^2}{\partial^2h/\partial v_x^2}}{1+\frac12\frac{\partial^4h/\partial z^4}{(\partial^2h/\partial z^2)^2}}.\label{N0_can_diff}
\end{equation}
The second derivatives of $h$ are given by Eqs.~(\ref{p0_BEC1_sder1}), (\ref{p0_BEC1_sder3}), and the higher derivatives can be estimated by transforming the sums (\ref{log_sum_der6}) to integrals:
\begin{equation}
\frac{\partial^3h}{\partial z\partial v_x^2}
=\sum_{s=1}^\infty e^{sz}\left.\frac{\partial^2G(v_x,s/R)}{\partial v_x^2}\right|_{v_x=0}G^2\left(0,\frac{s}R\right)\approx44.1R+o(R),\qquad\frac{\partial^4h}{\partial z^4}=R^4J_3(0,Rz)+o(R).
\end{equation}
Numerically, $\frac12(\partial^4h/\partial z^4)/(\partial^2h/\partial z^2)^2=\frac12J_3(0,Rz)/J_1^2(0,Rz)$ is close to zero at $|Rz|\ll1$ and does not exceed $0.4$ at larger $|Rz|$, so we can approximate the denominator of Eq.~(\ref{N0_can_diff}) by unity to obtain
\begin{equation}
\bar{N}_0-\bar{N}_{0\mathrm{c}}\approx\frac{33.1R^2}{N}\left(\frac{T_\mathrm{c}}T\right)^{3/2}\approx 0.93N^{1/3}\left(\frac{T}{T_\mathrm{c}}\right)^{1/2}.\label{delta_N0}
\end{equation}
As seen, the shift of the maximum location due to momentum fixation is subleading with respect to the maximum location itself (\ref{p0_BEC1_n0}), and the relative shift $(\bar{N}_0-\bar{N}_{0\mathrm{c}})/\bar{N}_0\sim N^{-2/3}(T/T_\mathrm{c})^{1/2}$ vanishes in the large-$N$ and low-$T$ limits.

Thus the distribution function in canonical ensemble (\ref{p_c}) is calculated similarly to $p(N_0)$, as described in Eqs.~(\ref{p0_BEC1_st1})--(\ref{p0_BEC1_distr}). The difference is the absence of derivatives $\partial^2h/\partial v_x^2$ and assuming $\mathbf{Q}=0$, so the canonical ensemble counterpart of the resulting distribution function (\ref{p0_BEC1_distr}) is
\begin{equation}
p_\mathrm{c}(N_0)\sim\exp\left\{-\frac{(N_0-\bar{N}_{0\mathrm{c}})^2}{2R^2J_1(0,\alpha_{0\mathrm{c}})}\right\},\label{can_distr}
\end{equation}
where $\alpha_{0\mathrm{c}}$ is slightly shifted with respect to $\alpha_0$ taken at $Q=0$ [see Eq.~(\ref{p0_BEC1_max_cond}) at $\tilde{v}_x=0$]. As seen from Eq.~(\ref{delta_N0}), the distribution maximum is shifted insignificantly, however for the distribution width the situation can be drastically different. Namely, using Eq.~(\ref{p6}) we can note that the terms $R^2J_1(0,\alpha_0)$ and $R^2J_1(0,\alpha_{0\mathrm{c}})$, entering Eqs.~(\ref{p0_BEC1_distr}) and (\ref{can_distr}), are just the second derivatives $\partial^2h/\partial z^2$ taken at different maximum locations $\bar{N}_0$ and $\bar{N}_{0\mathrm{c}}$, respectively. Thus the relative difference of the distribution dispersion is
\begin{equation}
\frac{\left.\frac{\partial^2h}{\partial z^2}\right|_{z=\tilde{z}(\bar{N}_0)}-\left.\frac{\partial^2h}{\partial z^2}\right|_{z=\tilde{z}(\bar{N}_{0\mathrm{c}})}}{\left.\frac{\partial^2h}{\partial z^2}\right|_{z=\tilde{z}(\bar{N}_0)}}\approx\frac{\partial^3h/\partial z^3}{\partial^2h/\partial z^2}\frac{d\tilde{z}}{dN_0}(\bar{N}_0-\bar{N}_{0\mathrm{c}}).
\end{equation}
Estimating the third derivative by transforming the sum (\ref{log_sum_der6}) to integral as $\partial^3h/\partial z^3\approx R^3J_2(0,Rz)$, and using Eqs.~(\ref{z_der_saddle}), (\ref{delta_N0}), we obtain
\begin{equation}
\frac{\left.\frac{\partial^2h}{\partial z^2}\right|_{z=\tilde{z}(\bar{N}_0)}-\left.\frac{\partial^2h}{\partial z^2}\right|_{z=\tilde{z}(\bar{N}_{0\mathrm{c}})}}{\left.\frac{\partial^2h}{\partial z^2}\right|_{z=\tilde{z}(\bar{N}_0)}}\approx-33.1\frac{J_2(0,R\tilde{z}(\bar{N}_0))}{J_1^2(0,R\tilde{z}(\bar{N}_0))}\frac{R}{N}\left(\frac{T}{T_\mathrm{c}}\right)^{-3/2}\approx-0.33N^{-1/3}\left(\frac{T}{T_\mathrm{c}}\right)^{-1/2}.
\end{equation}
Here we have taken into account that $J_2(0,Rz)/J_1^2(0,Rz)\approx0.06$ at $|Rz|\ll1$. As seen, $p(N_0)$ is generally more narrow than $p_\mathrm{c}(N_0)$. The change of distribution width due to momentum fixation is relatively small if $T/T_\mathrm{c}\gg N^{-2/3}$. At very low temperatures, $T/T_\mathrm{c}\lesssim N^{-2/3}$, the difference of the distribution widths cannot be neglected. As demonstrated above, in this case the free energy changes significantly due to momentum fixation as well, and the saddle-point approximation itself becomes poorly applicable.